\documentclass[fleqn,usenatbib,onecolumn]{mnras}

\usepackage{newtxtext,newtxmath}
\usepackage{pifont}% http://ctan.org/pkg/pifont
\newcommand{\cmark}{\ding{51}}%
\newcommand{\xmark}{\ding{55}}%
\usepackage[T1]{fontenc}
\usepackage[normalem]{ulem}
\usepackage{multirow}
\usepackage{comment}
\usepackage{floatrow}
\usepackage{graphicx}
\usepackage[label font=bf,labelformat=simple]{subfig}
\usepackage{caption}
\usepackage{soul}
\usepackage{tabularray}
\UseTblrLibrary{booktabs}
\DeclareRobustCommand{\VAN}[3]{#2}
\let\VANthebibliography\thebibliography
\def\thebibliography{\DeclareRobustCommand{\VAN}[3]{##3}\VANthebibliography}

\usepackage{xcolor}
\definecolor{green}{RGB}{21, 200, 0}
\usepackage{graphicx}	% Including figure files
\usepackage{amsmath}	% Advanced maths commands
\usepackage{orcidlink}
\newcommand{\orcid}[1]{\orcidlink{#1}}
\title[Forecasts for anisotropic stress]{Model-independent forecasts for the cosmological anisotropic stress}

\author[Z. Sakr et al.]{
Ziad Sakr~\orcid{0000-0002-4823-3757}$^{1,2}$,\thanks{E-mail:ziad.sakr@net.usj.edu.lb}\thanks{E-mail:sakr@thphys.uni-heidelberg.de}
Ziyang Zheng~\orcid{0000-0003-4396-059X}$^{1}$,\thanks{E-mail:Zheng@thphys.uni-heidelberg.de}
and Santiago Casas~\orcid{0000-0002-4751-5138}$^{3}$
\\
$^{1}$Faculty of Sciences, Universit\'e St Joseph, Beirut, Lebanon;\\
$^{2}$Institut f\"ur Theoretische Physik, University of Heidelberg, Philosophenweg 16, 69120 Heidelberg, Germany;\\
$^{3}$Institute for Theoretical Particle Physics and Cosmology (TTK), RWTH Aachen University, 52056 Aachen, Germany\\
}

\date{Accepted XXX. Received YYY; in original form ZZZ}

\pubyear{2025}

\begin{document}
\label{firstpage}
\pagerange{\pageref{firstpage}--\pageref{lastpage}}
\maketitle

\begin{abstract}
The effective anisotropic stress $\eta$ is a key variable in the characterization of many classes of modified gravity theories, as it allows the testing for a long-range force additional to gravity. In this paper we forecast the precision with which future large surveys can determine $\eta$ in a way that only relies on directly observable quantities obtained from the spectroscopic measurements of the clustering of galaxies and the photometric based observation of the projected lensing and galaxy clustering correlations and their cross signal. Our method does not require further assumptions about the initial power spectrum, the modified gravity model, the expansion rate, or
the bias. We consider various cases: $\eta$ free to vary in space and time, or with only redshift dependence, or constant. We take as a reference specifications that approximate a Euclid-like photometric or a combined one with a DESI-like spectroscopic survey. Among our results, we find that a future large-scale lensing and clustering survey can constrain $\eta$ to at least 30\% if $z$, $k$ independent, and to less than 10\% on average for the $z$ dependence only, to finally reach 5\% values in the constant case.
\end{abstract}


%%%%%%%%%%%%%%%%%%%%%%%%%%%%%%%%%%%%%%%%%%%%%%%%%%

%%%%%%%%%%%%%%%%% BODY OF PAPER %%%%%%%%%%%%%%%%%%
\section{Introduction}

With the latest constraints on the cosmological parameters from the cosmic microwave background (CMB) correlation measurements from the Planck satellite \citep{Planck:2018vyg}  and the recent DES \citep{DES:2021wwk} and DESI \citep{DESI:2024mwx} results, we reached the era of precision cosmology, in which most of the standard cosmological $\Lambda$CDM parameters are determined to percent level accuracy. However, the physical nature of the dark sector is completely unknown, and especially the cosmological constant suffers from severe theoretical problems. For this reason, it is of crucial importance to look beyond the perfectly homogeneous cosmological constant and to investigate general dark energy models, including also modifications of Einstein's theory of general relativity (GR). This is also allowed because current and upcoming cosmological surveys will reach a sensitivity that will afford us to test modifications of gravity at cosmological scales and possibly to distinguish them from standard scenarios \citep{Martinelli:2021hir}. These tests require  to use observations that probe the evolution of the background of the Universe and the formation of large-scale structures that result from the growth of primordial perturbations. 
  
In general, the extensions of the $\Lambda$CDM model that affects the evolution of the homogenous background of the Universe can be encapsulated in the normalized Hubble parameter, $E(z) = H(z)/H(z=0)$; while at linear perturbation level, a modification of scalar perturbations with respect to the $\Lambda$CDM model can be described by two functions: the first, denoted by $\mu (t, k)$,  modifies the standard Poisson equation and the second, $\eta(t,k)$, is the ratio of the two linear gravitational potentials $\Psi$ and $\Phi$ which enter the spatial and temporal part, respectively, of the perturbed Friedmann-Robertson-Walker (FRW) metric. For a non-relativistic perfect fluid, the effective anisotropic stress $\eta$ is not sourced by matter at the linear level, so it can be considered as a genuine indicator of modified gravity and a key variable to test for a non-minimal coupling of matter to gravity. 
The main difference of this work from previous ones is the fact
that we strive to reach a high level of model independence to avoid introducing a theoretical bias into the results.
In particular, as will become clear in the following, we do not need to specify a shape of the power spectrum, nor specific functional forms of the expansion rate, the growth rate, and the linear bias.

There have been several attempts to constrain or forecast the parameters $\eta$ and $\mu$, with different degrees of model independence. 
Studies using the CMB angular power spectrum such as \cite{Planck:2018vyg} or \cite{Sakr:2021ylx}, provided constraints on $\eta$ along with $\mu$ as free parameters. More recently \cite{Sakr:2023xnw}, using a combination of CMB and other probes, obtained bounds on $\eta$ but with the growth index, and its specific parameterisation, instead of $\mu$ as the other perturbation related free parameter and assuming a dynamical dark energy model, while the first three years of observations of the Dark Energy Survey \citep{DES:2022ccp} reported constraints on $\mu$ and $\Sigma$ where the latter quantity could be translated into $\eta$ through $\Sigma= \dfrac{\mu}{2}(1+\eta)$. This parameterization has also been used to obtain forecast constraints for upcoming experiments: e.g. \cite{Casas:2022vik} or \cite{Albuquerque:2024} exploited upcoming surveys such as Euclid \citep{Euclid:2024yrr} or the spectroscopic and continuum observables from the Square Kilometer Array Observatory (SKAO) \citep{SKA:2018ckk}. 
In \cite{Raveri:2021dbu} the authors used principal component analysis methods to constrain $\mu$ and $\eta$ in each redshift bin separately using multiple cosmological probes. However, 
they assumed a fixed shape for the power spectrum entering the $f\sigma_8(z)$ and assumed different parametrisations to model the bias for each measurements.
More recently \cite{Tutusaus:2023aux} produced forecasts by combining gravitational lensing and gravitational redshift to measure anisotropic stress with future galaxy surveys. 

 Assuming gravity remains universally coupled also when modified, one can build \citep{Amendola:2012ky} an estimate of $\eta(k,z)$ formed by three directly observable functions of scale and redshift that depend on the cosmic expansion rate, on the linear growth rate, and on the lensing correlation.
Euclid forecasts for this estimator have already been obtained in \cite{Amendola:2013qna}. In \cite{Pinho:2018unz} the same method has been applied to real data, but due to the lack of sufficient data, only very weak constraints on $\eta$ have been obtained. For the same reason, only a redshift-dependent $\eta$ has been considered.

The main aim of this paper is to improve upon the forecasts of \cite{Amendola:2013qna} in several directions.  First, we include several nuisance parameters due to intrinsic alignment or the Doppler shift associated with the random peculiar velocities of galaxies. Second, we update the survey with the most recent specifications from Euclid and include DESI, so as to cover a larger redshift range. Third, we include the photometric projected galaxy galaxy clustering as well as its cross-correlation signal with shear lensing.

% This is a simple template for authors to write new MNRAS papers.
% See \texttt{mnras\_sample.tex} for a more complex example, and \texttt{mnras\_guide.tex}
% for a full user guide.

\section{Theory and Methods}\label{sec:theoandmeth}

\subsection{Four model-independent quantities}
\label{sec:maths} % used for referring to this section from elsewhere
We begin with a perturbed flat Friedmann-Lemaître-Robertson-Walker (FLRW)  metric, considering only scalar perturbations in the Newtonian gauge,
\begin{equation}
\mathrm{d}s^{2}=-a^{2}(1+2\Psi)\mathrm{d}\tau^{2}+a^{2}(1+2\Phi)\mathrm{d}x_{i}\mathrm{d}x^{i} \,,
\end{equation}
where $\Psi$ and $\Phi$ are the two gravitational potentials. Hereafter in this paper we adopt Planck units, i.e.  $c = G_{\rm N} =1$. Assuming a pressureless perfect fluid for matter and a flat Universe, one can derive the two gravitational potential equations \citep{Amendola:2019laa} that relate $\Psi$ and $\Psi$ to the matter density distribution,
\begin{align}\label{eq: Psi}
k^{2}\Psi&=-4\pi a^{2}\mu(k,z)\bar{\rho}_m(z)\delta_m(k,z)  \,,
\\
k^{2}\Phi&= 4\pi a^{2}\mu(k,z)\eta(k,z)\bar{\rho}_m(z)\delta_m(k,z)  \,,
\end{align}
where $\bar{\rho}_m$ is the average background matter density, $k$ is the comoving wavevector, and $\delta_m=\frac{\delta\rho_m}{\bar{\rho}_m}$ is the root-mean-square matter density contrast. $\eta$ and $\mu$ are two functions quantifying modified gravity.  In General Relativity they reduce to $\mu = 1$ and $\eta =1$, respectively. The linear anisotropic stress, $\eta$ can then be extracted by taking the ratio of the two Poisson equations:
\begin{equation}
\eta (k,z) =-\frac{\Phi}{\Psi}  \,.
\end{equation}
Notice that everywhere in this paper the perturbation variables represent root-mean-square quantities, so are positive definite.
Substituting $\bar{\rho}_m$ with the fractional matter density $3H^2\Omega_m(z)/8\pi$, Eq. \eqref{eq: Psi} becomes
\begin{equation}\label{eq:Poisson}
k^{2}\Psi=-\frac{3}{2}\mathcal{H}^{2}(z)\Omega_{{\rm m}}(z)\mu(k,z)\delta_m(k,z)  \,,
\end{equation}
where $\mathcal{H}=aH$. 

The evolution equation for linear matter perturbations in a generalized gravity theory with modified gravity parameter $\mu(k,z)$ is given by
\begin{equation}\label{eq:linear}
\delta_m''+ \delta_m'\left(2+\frac{E'}{E}\right)=-\left(\frac{k}{\mathcal{H}}\right)^2\Psi \,,
\end{equation}
in which we use a prime to denote a derivative with respect to the $e$-folding time $N=\ln a$. Expressing Eq. \eqref{eq:linear} in terms of the growth
rate $f=\delta_m'/\delta_m$ and inserting Eq. \eqref{eq:Poisson}, we have 
\begin{align}\label{eq:f-1}
f'+f^{2}+\left(2+\frac{E'}{E}\right)f & =\frac{3}{2}\Omega_{{\rm m}}(z)\mu(k,z)\,.
\end{align}
As pointed out in \cite{Amendola:2013qna}, cosmological observations
at large (linear) scales can measure three model-independent quantities.
 Besides the dimensionless expansion rate $E(z)\equiv H(z)/H_{0}$,
these are the galaxy power spectrum amplitude, the redshift distortion
amplitude and the weak shear lensing amplitude, defined respectively
in Fourier space as 
\begin{align}
\label{eq:mod-ind_quan}
A(k,z) & =G(k,z)b(k,z)\sigma_8\delta_{\rm m,0}(k)\,, \\ R(k,z)&=G(k,z)f(k,z)\sigma_8\delta_{\text{m,0}}(k),\\
L(k,z) & =\Omega_{{\rm m},0}\mu(k,z)[1+\eta(k,z)]G(k,z)\sigma_8\delta_{\text{m,0}}(k)\,.
\end{align}
where $G(k,z)=\delta_m(k,z)/\delta_{\rm m,0}(k)$ is the normalized growth, $f(k,z)=\delta_m'(k,z)/\delta_m(k,z)$,
$b(k,z)$ is the linear bias, and $\delta_{\rm m,0}(k)$ is the present
square root of the matter power spectrum normalized with the variance in cells with radius 8 Mpc$/h$,
$\sigma_{8}$. All the $A,R,L,E$ parameters are positive definite.

We take now suitable combinations
of the above observable quantities:
\begin{align}
P_{1} & \equiv  RA^{-1}=f/b,\\
P_{2} & \equiv  LR^{-1}=\Omega_{{\rm m},0}\mu(1+\eta)/f,\\
P_{3} & \equiv  d\log R/d\log a=f+f'/f
\end{align}
 and, by combining with the standard evolution equation (\ref{eq:f-1}),
since $\Omega_{m}=\Omega_{{\rm m},0}(1+z)^{3}/E^{2}$, we obtain
the relation
\begin{equation}
\frac{3P_{2}(1+z)^{3}}{2E^{2}\left(P_{3}+2+\frac{E'}{E}\right)}-1=\eta\,\,.\label{eq:eta}
\end{equation}
from which also $\Omega_{{\rm m},0}$ and $b$ are finally also absent. In
this sense, Eq.~\ref{eq:eta} is a model-independent test
of gravity \footnote{Note the $P_{2}$ is related to the $E_{G}$ statistics (see the recent study by \cite{Li:2025mib} and references therein), whose value at a scale $k$ is as $E_{g}=\left\langle \frac{a\nabla^{2}(\Psi-\Phi)}{3H_{0}^{2}f\delta_m}\right\rangle_{k}$. In our definitions, the relation would then be $P_{2}=2E_{g}$.}. 

\subsection{Galaxy spectroscopic power spectrum and the 3$\times$2pt joint analysis of photometric weak lensing and galaxy clustering}\label{sec:photo_spectro_obs}
\subsubsection{Galaxy power spectrum}
The observed linear galaxy power spectrum can be written as  
\begin{equation}
P_{\rm obs}(k,\mu, z)=G^{2}(k,z)b^{2}(k,z)(1+u^{2}\frac{f}{b})^{2}\sigma_8^2\delta^2_{\rm m,0}(k)e^{-k_\parallel^{2}u^{2}\sigma_{r}^{2}}\left\{\frac{1}{1+[k\,u\ \sigma_\text{p}(z)]^2}\right\}\label{gcpower}
\end{equation}
where $\sigma_{r}=\sigma_{0,z}(1+z)/H(z)$, $\sigma_{0,z}$ being the absolute error on redshift measurement, noting that the damping due to redshift errors does not vary with changes in the expansion history since $k_\parallel \propto H(z)$ and $\sigma_r \propto H^{-1}(z)$, and $u$ is the cosine of the angle between the line of sight and the wavevector, while the last term in  the curly brackets is a Lorentzian contribution, accounting for the Finger-of-God effect with $\sigma_{\rm p}(z)$ being the galaxy velocity dispersion\footnote{We note that our analysis being independent of the detailed power spectrum shape, the latter is assumed not strongly sensitive to the exact location of the Baryonic Acoustic Oscillations (BAO) wiggles \citep{Amendola:2022vte}, and therefore of the effect of bulk flows on them, which translates in a damping factor on the oscillating part of the power spectrum. We do not include this effect here, since it could also serve as additional constraints on our parameters, while it was recommended to remain a nuisance effect \citep{Wang:2012bx}, and we limit ourselves to only including BAO as part of our parameters that include the power spectrum as one of their ingredients}. In our parameters the observed power spectrum will then be :
\begin{equation}
P_{\rm obs}(k,\mu, z)=(A+Ru^{2})^{2}e^{-k^{2}u^{2}\sigma_{r}^{2}} \left\{\frac{1}{1+[k\,u\ \sigma_\text{p}(z_i)]^2}\right\},\label{gcpowerAR}
\end{equation}
where $\sigma_{\rm p}(z)$ are nuisance free parameters at each of the same redshift bins division we choose for our model independent parameters and  
The dependence on $E=H/H_{0}$ is implicitly contained in $u$ and
$k$ through the Alcock-Paczynski effect. Explicitly,
$u,k$ depend on the fiducial $u_{f}, k_f$ (hereafter we use subscript $f$ to denote quantities at the fiducial) via the relation 
\begin{align}\label{eq:mu}
u & =u_{f}\left[u_{f}^{2}-\frac{E_{f}^{2}{D_A}_{f}^{2}}{E^{2}{D_A}^{2}}(u_{f}^{2}-1)\right]^{-1/2} \\
k & =k_{f}\frac{E}{E_f}\left[u_{f}^{2}-\frac{E_{f}^{2}{D_A}_{f}^{2}}{E^{2}{D_A}^{2}}(u_{f}^{2}-1)\right]^{1/2} 
\end{align}
where ${D_A}$ is the dimensionless angular diameter distance. In a spatially flat Universe, an assumption we adopt in this work, it reads:
\begin{equation}\label{eq:DA}
{D_A}=\frac{1}{(1+z)}\int_{0}^{z}\frac{dz'}{E(z')} \,.
\end{equation}

We leave $E$ to vary in our Fisher implementation in all places where it is explicitly or implicitly contained. For example, $E$ is varied in $\sigma_r$ and in Eq. \eqref{eq:mu} - Eq. \eqref{eq:DA}.

\subsubsection{Photometric lensing and galaxy auto- and cross-correlation probe}
For weak lensing, the observed  angular lensing-lensing convergence
power spectrum from a survey divided into several redshift bins can be expressed as \citep{DES:2022ccp} 
\begin{equation}
    C_{ij}^{\rm \gamma\gamma}(\ell) = \int_0^\infty{\rm d}z\,\frac{W_i^{\rm \gamma}(z)W_j^{\rm \gamma}(z)}{H(z)r^2(z)}\left[\frac{\mu}{2}(1+\eta)\right]^2 P_{\delta_{\rm m}\delta_{\rm m}}^{\rm }(k,z)\,,\label{eq:CL_WL}
\end{equation}
where $P_{\delta_{\rm m}\delta_{\rm m}}(k,z)$ is the matter power spectrum evaluated at $k=k_{\ell}(z)=\frac{\ell + 1/2}{r(z)}$,
and $i$ and $j$ denote two tomographic redshift bins. The lensing weights $W_i^{\rm \gamma}(k,z)$ are given by:
\begin{align}
W_i^{\rm \gamma}(k,z) =&\; \frac{3}{2}\Omega_{{\rm m},0} H_0^2(1+z)r(z) \,
 \int_z^{z_{\rm max}}{dz'{n_i(z')}\frac{r(z')-r(z)}{r(z')}} \, ,
\end{align}
where $n_i(z)$ is the normalised redshift distribution of galaxies in the $i$-th bin \citep{Euclid:2019clj}. 
Note that $E$ is implicitly contained in the comoving distance $r(z)$, here and in any of the subsequent equations where it figures.

Writing Eq. (\ref{eq:CL_WL}) as a function of the above defined model-independent quantities from Eq. (\ref{eq:mod-ind_quan}), it becomes:
\begin{equation}
    C_{ij}^{\rm \gamma\gamma}(\ell)= \int{\rm d}z\,\frac{K_i^{\rm \gamma}(z)K_j^{\rm \gamma}(z)}{E(z)}\frac{1}{4}L^2{\rm }\left(\frac{\ell+1/2}{r(z)},z\right)\,
\end{equation}
where 
\begin{align}
K_i^{\rm \gamma}(k,z) =&\; \frac{3}{2} H_0^2(1+z) \,
 \int_z^{z_{\rm max}}{dz'n_i(z')\frac{r(z')-r(z)}{r(z')}} \, . 
\end{align}
We should add to the previous and subsequent lensing quantities a shot noise component from the uncorrelated part of the intrinsic (unlensed) ellipticity field that can be written as 
\begin{equation}
N^\epsilon_{ij}(\ell)=\frac{\sigma^2_\epsilon}{n_i}\delta^{\rm K}_{ij} \,,
\end{equation}
where $n_{i}$ is the galaxy surface density in the  bin $i$, $\delta^{\rm K}_{ij}$ is the Kronecker delta symbol; and $\sigma^2_\epsilon$ is the variance of the observed ellipticities.

We also included intrinsic alignment (IA) effects into our formalism, where the correlation between background shear and foreground intrinsic alignment $C^{\rm I\gamma}_{ij}(\ell)$, and the autocorrelation of the foreground intrinsic alignment $C^{\rm II}_{ij}(\ell)$, are given, respectively, by
\begin{align}\label{eq:newmod_IA_Cell}
C^{\rm I\gamma}_{ij}(\ell) & = \int\mathrm d z\,{\frac{W_{i}^{\gamma}(z) W_{j}^{\rm IA}(z) + W_{i}^{\rm IA}(z) W_{j}^{\gamma}(z)}{H(z) r^2(z)} \frac{\mu}{2}(1+\eta) P_{\delta_{\rm m}\delta_{\rm I}}\!\left[ \frac{\ell + 1/2}{r(z)}, z \right]}, \nonumber\\
C^{\rm II}_{ij}(\ell) & =\int\mathrm d z\,{\frac{W_{i}^{\rm IA}(z) W_{j}^{\rm IA}(z)}{H(z) r^2(z)} P_\mathrm{\delta_{\rm I}\delta_{\rm I}}\!\left[ \frac{\ell + 1/2}{r(z)}, z \right]}.
\end{align}
where the corresponding weight function are expressed as 
\begin{equation}
W_i^{\rm IA}(z) =  \frac{n_{i}(z)}{1/H(z)} = H_0 \, n_i(z) E(z) \,.
\label {eq:iaweight}
\end{equation}
 and $P_{\delta_I\delta_m}$ and $P_{\delta_I\delta_I}$ are the power spectra relative respectively to $\delta_m$ auto and cross correlations, and $\delta_I$ the intrinsic alignment density contrast, related to the matter density one as \citep{Troxel:2014dba}
\begin{equation}
    \delta_{\rm I} = -{\cal{A}}_{\rm IA} {\cal{C}}_{\rm IA} \, \mu(k,z)\,\Omega_{{\rm m},0}\frac{{\cal{F}}_{\rm IA}(z)}{G(z,k)} \delta_{\rm m}(k,z)
    \label{eq:pdienla}      \,,
\end{equation}
  where we see that the factor $\mu$, an ingredient of one of our parameters, was introduced since in this IA formalism, Eq.~\ref{eq:pdienla} results essentially from a Poisson potential equation \citep{Hirata:2004gc} (see the appendix \ref{sec:IA_model} for more details). 
Note that we also divide by the growth in which $\mu$ is also absorbed as being part of the commonly used sub-horizon growth equation see e.g. \citep{Zheng:2023yco},
 and as is the case for similar quantities in Eq.~\ref{gcpowerAR} and Eq.~\ref{eq:CL_WL}. It remains the quantity ${\cal{F}}_{\rm IA}(z)$ which is equal to $(1 + z)^{\eta_{\rm IA}}[\langle L_g \rangle (z)/ L_{g\star}  (z)]^{\beta_{\rm IA}}$ \citep{Euclid:2019clj}, where $\langle L_g(z)\rangle$ and $ L_{g\star}(z)$ are the redshift-dependent mean and the characteristic luminosity of source galaxies, respectively, as computed from the luminosity function. $\eta_{\rm IA}$ and $\beta_{\rm IA}$ are the redshift and power law dependence parameters of the luminosity function while $\mathcal{A}_{\rm IA}$ and $\mathcal{C}_{\rm IA}$ are further constant nuisance parameters. We leave $\mathcal{A}_{\rm IA}$, $\beta_{\rm IA}$ and $\eta_{\rm IA}$ free to vary,  and 
  fix $\mathcal{C}_{\rm IA}$ as it is degenerate with $\mathcal{A}_{\rm IA}$.  
At the end, as function of our above defined quantities, the intrinsic alignment and lensing equations become:
\begin{align}
C^{\rm I\gamma}_{ij}(\ell) & = \int{\rm d}z\,\frac{K_i^{\rm I}(z)K_j^{\rm \gamma}(z)}{E(z)\,r(z)}\frac{1}{2(1+\eta)}L^2{\rm }\left(\frac{\ell+1/2}{r(z)},z\right)\, \nonumber\\
C^{\rm II}_{ij}(\ell) & =\int{\rm d}z\,\frac{K_i^{\rm I}(z)K_j^{\rm I}(z)}{E(z)\,r(z)^2}\frac{1}{4(1+\eta)^2}L^2{\rm }\left(\frac{\ell+1/2}{r(z)},z\right)\,,
\end{align}
where 
\begin{align}
K_i^{\rm I}(k,z) = -{\cal{A}}_{\rm IA} {\cal{C}}_{\rm IA} \, \frac{L(z,k)}{L(0,k)} \, H_0 \, n_i(z) E(z).
\end{align}

Finally, we also include the photometrically detected galaxy-galaxy correlations, with the radial weight function for galaxy clustering defined as
\begin{equation}\label{eq:photoGCwin}
W^{\rm G}_i (k,z)=b_i(k,z) n_i (z){H(z)} \,= b_i(k,z) n_i (z) H_0  E(z),
\end{equation}
where $b_i(k,z)$ is the galaxy bias in the $i$-th redshift bin. We multiply by $\delta(k,z)$,
to obtain the galaxy-galaxy autocorrelation or the galaxy-galaxy lensing cross correlations.
The factor $b(k,z)\delta(k,z)$ would then be replaced by  $A(k,z)$ assuming same bias for the spectroscopic- and photometric-detected galaxies: 
\begin{equation}
    C_{ij}^{\rm GG}(\ell)= \int{\rm d}z\,\frac{K_i^{\rm G}(z)K_j^{\rm G}(z)}{E(z) r(z)^2}\frac{1}{2}A^2{\rm }\left(\frac{\ell+1/2}{r(z)},z\right)\,
\end{equation}
where 
\begin{align}
K_i^{\rm G}(k,z) = n_i (z) \, H_0 \, E(z), 
\end{align}

The same formalism is extended to additionally include in the analysis the cross correlation between galaxy and galaxy lensing (or the intrinsic alignment alike signal) given by:
\begin{equation}
    C_{ij}^{XY}(\ell)=\int_{z_{\rm min}}^{z_{\rm max}}\text{d}z\frac{W_i^{\rm X}(z)\,W_j^{\rm Y}(z)}{H(z)\,r^2(z)}P_{\delta_{\rm A}\delta_{\rm B}}(k_{\ell},z),
\end{equation}
where $i$ and $j$ refer to two tomographic redshift bins, $\rm X$ and $\rm Y$ stand for either the clustering or the lensing probe, and $\rm A$ and $\rm B$ for $\rm m$ (matter) or $\rm I$ (intrinsic). For instance, if we use our model independent parameters one combination could be written as:
\begin{equation}
    C_{ij}^{G\gamma}(\ell)=\int{\rm d}z\,\frac{K_i^{\rm G}(z)K_j^{\rm \gamma}(z)}{E(z)\,r(z)}\sqrt{\frac{1}{2}}A{\rm }\left(\frac{\ell+1/2}{r(z)},z\right)\sqrt{\frac{1}{4}}L{\rm }\left(\frac{\ell+1/2}{r(z)},z\right),
\end{equation}

\subsection{Fisher matrix formalism and datasets}
\label{sec:fisher}

\subsubsection{Settings}

 For the spectroscopic survey, we join a DESI-like survey at low redshift to a Euclid-like one at higher redshift, according to Table~\ref{tab:Surveys_spec}.
The DESI-like survey reproduces the specifications for the DESI Bright Galaxy Survey for $z\le 0.6$ based on \cite{Hahn:2022dnf}, and the DESI Emission Line Galaxies (ELG) survey for $0.6\le z \le 0.9$ based on \cite{DESI:2016fyo}, while for $0.9\le z \le 1.7$ we assume a Euclid-like survey based on \cite{Euclid:2019clj}. We call this the DE combined survey.  For the photometric survey, we also assume a Euclid-like settings as shown in Table~\ref{tab:Surveys_spec} following \cite{Euclid:2019clj} but adopt equi-spaced bins in which the $n(z)$ are interpolated from the ones in the equi-populated bins in the referred study.  

As already mentioned, we leave our parameters $A,R,L,E$ free to vary in every redshift and $k$ bin. So the first task is to define these bins.
The expansion rate $E(z_i)$ is divided into six bins of size $\Delta z=0.2$ centred on  
\begin{equation}\label{eq:modelzbins}
z_{i} = \{0.6,0.8,1.0,1.2,1.4,1.6\},
\end{equation}
Moreover, for the quantities that depend on $k$, namely $L,R$ and $A$, we take four $k$ bins  which central values 
\begin{equation}
k=\{0.0075,0.03,0.075,0.125\}
\end{equation}
and corresponding boundaries 
\begin{equation}
k=\{0.005,0.01,0.05,0.1,0.15\}
\end{equation}
 so that the number of parameters for each quantity $L, R$ or $A$ is $6\times4 = 24$. 

We choose the following fiducial values of $\Lambda$CDM:
\begin{equation}
 \Omega_{\rm m,0}  = 0.315\,,\quad
   \Omega_{\rm b,0}  = 0.049\,,\quad
 h  = 0.6737\,,\quad  n_{\rm s}  = 0.96\,,\quad   \sigma_8  = 0.81,
\end{equation} 
and adopt for the fiducial bias the function $b = \sqrt{1+z}$ from \cite{Clerkin:2014pea} and compute using the linear matter power spectrum $\sigma_\text{p}^2(z_i)=\frac{1}{6\pi^2}\int P_{\delta \delta} (k,z_i)\;{\rm d}k\ $ \citep{Euclid:2019clj}.  For the lensing nuisance parameters we adopt the values from \cite{Euclid:2019clj} $\{\mathcal A_{\rm IA},\eta_{\rm IA},\beta_{\rm IA},\mathcal{C}_{IA}\}$ $=\{1.72,-0.41,2.17,0.0134\}$. We use our $A$, $R$, $L$ and $E$ binned parameters to construct an interpolator following a cubic spline method and use it to obtain the values of the relevant quantities at the desired redshift and wave-number.
\begin{table}
\resizebox{0.7\columnwidth}{!}{%
%\begin{tabular}{|c | c |} 
\begin{tabular}{l  c c c c c c c c c c}  
\hline
\multirow{2}{*}{cosmo. param.} & $\Omega_{\rm b,0}$	& $n_s$ & {$\Omega_{\rm m,0} $} & {$h $} & {$\sigma_8$}\\
 & 0.049 & 0.96 & 0.315 & 0.6732 & 0.81   \\ \hline
\multirow{2}{*}{nuis. param.} & $\mathcal A_{\rm IA}$ & $\eta_{\rm IA}$ & $\beta_{\rm IA}$ & $\mathcal{C}_{\rm IA}$ & $\sigma_{\rm p}(z_1)$ & $\sigma_{\rm p}(z_2)$ & $\sigma_{\rm p}(z_3)$ & $\sigma_{\rm p}(z_4)$ & $\sigma_{\rm p}(z_5)$ & $\sigma_{\rm p}(z_6)$   \\
 & 1.72 & -0.41 & 2.17 & 0.0134 & 4.484 & 4.325 & 4.121 & 3.902 & 3.683 & 3.475 \\ \hline
\end{tabular}
}
\caption{Cosmological and nuisance parameter fiducial values as adopted in \citet{Euclid:2019clj}}. 
\label{tab:baseline_parameters}
\end{table}

\renewcommand{\arraystretch}{1.7}
\begin{table}
\centering
\resizebox{\columnwidth}{!}{
%\begin{tabular}{|c | c c c c c|} 
\begin{tabular}{l  c c c c c c} 
\hline
\multirow{2}{*}{Euclid 3$\times$2$pt$ $C_{\ell s}$ photo} & $A_{\rm surv}({\rm deg}^2)$ & $z_{{\rm obs},i} ({\rm edges})$ & $\bar n_{\rm gal}(\mathrm{arcmin}^{-2})$ & $\sigma_\epsilon$ & $\ell_{\rm min}$    \\
 & 15000 & $\{0.001, 0.1, 0.3, 0.5, 0.7, 0.9, 1.1, 1.3, 1.5, 1.9, 2.3\}$ & 30 & 0.3 &  10  \\ \hline
\multirow{2}{*}{Euclid+DESI $P_{\rm k}$ spectro} & $z_{{\rm obs},i} ({\rm edges})$ & $\bar n_{{\rm gal},i} \, (h^3 {\rm Mpc}^{-3})$ & $V_i \, (h^{-3} {\rm Gpc}^{3})$ & $\sigma_{0,z}$ & $k_{\rm min}(h\,{\rm Mpc}^{-1})$ & $k_{\rm max}(h\,{\rm Mpc}^{-1})$    \\
 & $\{0.5, 0.7, 0.9, 1.1, 1.3, 1.5, 1.7\}$ & $\{2.03, \,9.57,\,6.82,\,5.54,\,4.18,\,2.62\}$ $\times 10^{-4}$ & $\{4.56, \,6.42,\, 7.98,\, 9.20,\, 10.11,\, 10.77\}$ & 0.001 &  0.005 &  0.15  \\ \hline
\end{tabular}
}
\caption{Euclid-like photometric angular and Euclid + DESI-like spectroscopic survey 3D power spectrum specifications taken from \citet{Euclid:2019clj}, \citet{DESI:2016fyo} and \citet{Hahn:2022dnf}, with $A_{surv}$ the survey area, $V_i$ the survey volume in each redshift bin, $\sigma_\epsilon$ the intrinsic ellipticity dispersion, and $\sigma_{0,z}$ the error on the photometric redshift measurement}
\label{tab:Surveys_spec}
\end{table}

\subsubsection{Fisher matrix} 
The Fisher matrix for the clustering probe from spectroscopic measurements, for a parameter vector $p_\alpha$, is generally given by \begin{equation}
F_{\alpha\beta}^{\text{GC}}(z_i)=\frac{1}{8\pi^{2}}\int_{-1}^{1}{\mathrm{d}u}\int_{k_{\text{min}}}^{k_{\text{max}}}k^{2}V_{\text{eff}}\frac{d\log P}{dp_{\alpha}}\biggl|_{f}\frac{d\log P}{dp_{\beta}}\biggl|_{f}\,{\mathrm{d}k}\,,\label{eq:fmgc}
\end{equation}
where the effective survey volume, $V_{\text{eff}}$, is
\begin{equation}
V_{\text{eff}}(k,\mu; z)=\left(\frac{\bar n_{{\rm gal},i}(z)P_{\rm obs}(k,u;z)}{\bar n_{{\rm gal},i}(z)P_{\rm obs}(k,u;z)+1}\right)^{2}V_{\text{i}}(z)\label{veff} \,,
\end{equation}
where $V_i$ is the redshift bin volume, $\bar n_{{\rm gal},i}$ the galaxy number density in each bin and $P_{\rm obs}$ calculated at the fiducial.

Our parameter vector for the galaxy clustering probe is
\begin{equation}
p_{\alpha}^{\text{\tiny{GC}}}=\{A(z_i, k_j), R_{\rm GC}(z_i,k_j), E_{\rm GC}(z_i)\} \,,
\end{equation}
where the subscripts $i$ and $j$ run over the $z$ and $k$ bins, respectively. Greek indices label the parameters in the Fisher matrix, which is
evaluated at the fiducial, assuming a scale-independent fiducial bias in $\Lambda$CDM. In addition, the $k$ and $u$ integrations in Eq. (\ref{eq:fmgc}) are performed numerically using a trapezoidal double integration method, with the integrand represented as a matrix indexed by $u$ and $k$. The derivatives are then calculated following the three-point stencil numerical method where a spline interpolation reconstruction of our vector of parameters is applied to obtain the values of two points around the fiducial with 5\% as the step of differentiation.

The combined Fisher matrix for survey of photometric galaxy clustering, weak lensing, and their cross-correlation, that covers a fraction of the sky $f_{{\rm sky}}$, is a sum over $\ell$ bins \cite[see e.g.][]{Euclid:2019clj}
\begin{equation}
F_{\alpha\beta}^{\text{XC}}=\frac{1}{2}\sum_{\ell = \ell_{\rm min}}^{\ell_{\rm max}}(2\ell+1)\sum_{ABCD}\sum_{ij,mn}\frac{ C^{AB}_{ij}}{\partial p_{\alpha}}\left[\Delta C^{-1}(\ell)\right]^{BC}_{jm}\frac{ C^{CD}_{mn}}{\partial p_{\beta}}\left[\Delta C^{-1}(\ell)\right]^{DA}_{ni}\,,\label{eq:wlfm}
\end{equation}
where the block descriptors $A,B,C,D$ run
over the combined probes lensing and clustering and the indices $i,j,m,n$ are implicitly summed over, while 
\begin{align}
             \Delta{C}^{AB}_{ij}(\ell) &= \frac{1}{\sqrt{f_{\rm sky} \Delta \ell}}\left[C^{AB}_{ij}(\ell) + N^{AB}_{ij}(\ell)\right],
             \label{eq: covsecond}
       \end{align} 
       with $f_{\rm sky}$ the fraction of the sky obtained from $A_{\rm surv}$ in table~\ref{tab:Surveys_spec}. Here the parameters are $p_{\alpha}$=$\{A(z_1, k_1), R(z_1,k_1), L (z_1,k_1),E(z_1),\dots\, \alpha_{\rm IA}, \beta_{\rm IA},\gamma_{\rm IA}\}$, while $\ell$ is being summed from $\ell_{{\rm min}} = 10$ to $\ell_{{\rm max}}(z)= k_{\rm max}  r(z) -1/2$, where $k_{\rm max} = 0.125 \, h \rm {Mpc}^{-1}$ with $\Delta\ln\ell=0.1$. 

After marginalising over the nuisance parameters, the total Fisher matrix is obtained by summing the contributions from spectroscopic and photometric measurements for the common elements of $A, R$, and $E$. The full Fisher structure is given by
\begin{eqnarray}
 &  & \left(\begin{array}{cccc}
(AA)^{\Sigma} & (AR)^{\Sigma} & AL & (AE)^{\Sigma}\\
(RA)^{\Sigma} & (RR)^{\Sigma} & RL & (RE)^{\Sigma}\\
LA & LR & LL & LE\\
(EA)^{\Sigma} & (ER)^{\Sigma} & EL & (EE)^{\Sigma}
\end{array}\right) \,,
\end{eqnarray}
where $(\mathcal{K})^{\Sigma}=(\mathcal{K})^{\text{GC}}+(\mathcal{K})^{\text{WL$\times$ GCph}}$, and $\mathcal{K}=\{AA,AR,AE,RA,RR,RE,EA,ER,EE\}$ . We then marginalise over $A$ to obtain the Fisher matrix only on $R, L$ and $E$.

In our numerical approach, $R^\prime$ and $E^\prime$ are approximated as $R_i^\prime=(R(z_{i+1},k_j)-R(z_{i-1},k_j))/\Delta N_i$ and $E_i^\prime=(E(z_{i+1})-E(z_{i-1}))/\Delta N_i$, respectively, where $\Delta N_i=\ln[(1+z_{i-1})/(1+z_{i+1})]$.
Therefore, from Eq. \eqref{eq:eta}, the gravitational slip $\eta(z_i, k_j)$ can be evaluated at each $z,k$ bin as follows,
\begin{align}\label{eq:eta_num}
    \eta(z_i,k_j) = \frac{3\frac{L(z_i,k_j)}{R(z_i,k_j)}(1+z_i)^3}{2E_i^2\left[\frac{R(z_{i+1},k_j)-R(z_{i-1},k_j)}{\Delta N_iR(z_i,k_j)}+2+\frac{E(z_{i+1})-E(z_{i-1})}{\Delta N_iE(z_i)}\right]}-1\,,
\end{align}

We employ a similar Jacobian approach as in \cite{Zheng:2023yco} to evaluate the errors on $\eta$. Specifically, we assume that the distribution of $\eta$ is Gaussian and expand $\eta$ around the fiducials. The covariance matrix of $\eta$ is expressed as
\begin{align}
\begin{split}
&\sigma^2_{\eta(z_i,k_j)\eta(z_{i'}, k_{j'})}=\langle\left(\mathcal{\eta}(z_i,k_j)-1\right)\left(\mathcal{\eta}(z_{i'},k_{j'})-1\right)\rangle \\ =&\left\langle\left(\sum_{p=1}^{7}\frac{\partial \eta(z_i,k_j)}{\partial X_p^{(i,j)}}\biggr|_{\vec{X}^{(i,j)}_{(\mathcal{F})}}\Delta X_p^{(i,j)}\right)\left(\sum_{q=1}^{7}\frac{\partial \eta(z_{i'},k_{j'})}{\partial X_q^{(i',j')}}\biggr|_{\vec{X}^{(i',j')}_{(\mathcal{F})}}\Delta X_q^{(i',j')}\right)\right\rangle\\
=& \sum_{p=1}^{7}\sum_{q=1}^{7}\frac{\partial \eta(z_i,k_j)}{\partial X_p^{(i,j)}}\biggr|_{\vec{X}^{(i,j)}_{(\mathcal{F})}}\frac{\partial \eta(z_{i'},k_{j'})}{\partial X_q^{(i',j')}}\biggr|_{\vec{X}^{(i',j')}_{(\mathcal{F})}}\sigma^2_{X_p^{(i,j)}X_q^{(i',j')}}  \,.
\end{split}
\end{align}
Here we have defined  $\vec{X}^{(i,j)}=\{L(z_i,k_j), R(z_{i+1},k_j), R(z_i,k_j), R(z_{i-1},k_j),E(z_{i+1}), E(z_i),E(z_{i-1})\}$, and the subscript $(\mathcal{F})$ denotes the values at the fiducial. 

\begin{figure*}
\begin{center}
\includegraphics[width=3.6in,height=2.5in]{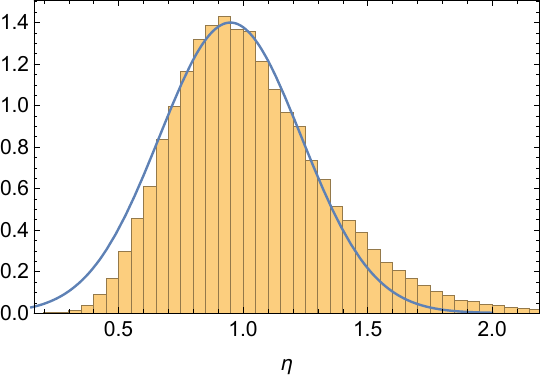}
\caption{Numerical distribution of $\eta$ for the wave-number $k=0.1 h/$Mpc, a value within the range of our $k$ bins. The blue curve is a Gaussian fit. } \label{fig:etagaussian}
\end{center}
\end{figure*}

We also tested the Gaussianity of $\eta$ by generating 30,000 values of $R,L,E$ for each bin, distributed as multi-Gaussian variables with a covariance matrix given by the inverse of the Fisher matrix (marginalized over the $A$'s parameters and the three nuisance parameters) and centred around the fiducial values. A typical distribution is shown in Fig. {\ref{fig:etagaussian}}, indicating that the Gaussian approximation is reasonably good.

% \textcolor{red}{(LA if you prefer not to use the numerical method, we could estimate the error from fig 1 and show quantitatively that indeed is similar to the Gaussian approx)(ZZ We sill need to decide to keep Fig 1 or not)
% }

% \textcolor{red}{(LA yes why not? in any case we should also estimate how good the error from the Gaussian in fig 1 approximates the numerical one.)(ZZ: The numerical $\eta$ in Fig. 1 is now obtained in a (k,z) bin that we do not use here, should we generate a new results, in order to compare to our results? )}

\section{Results and discussion}\label{sec:ResandDisc}

\begin{figure}[htbp]
\centering
\includegraphics[width=0.35\textwidth]{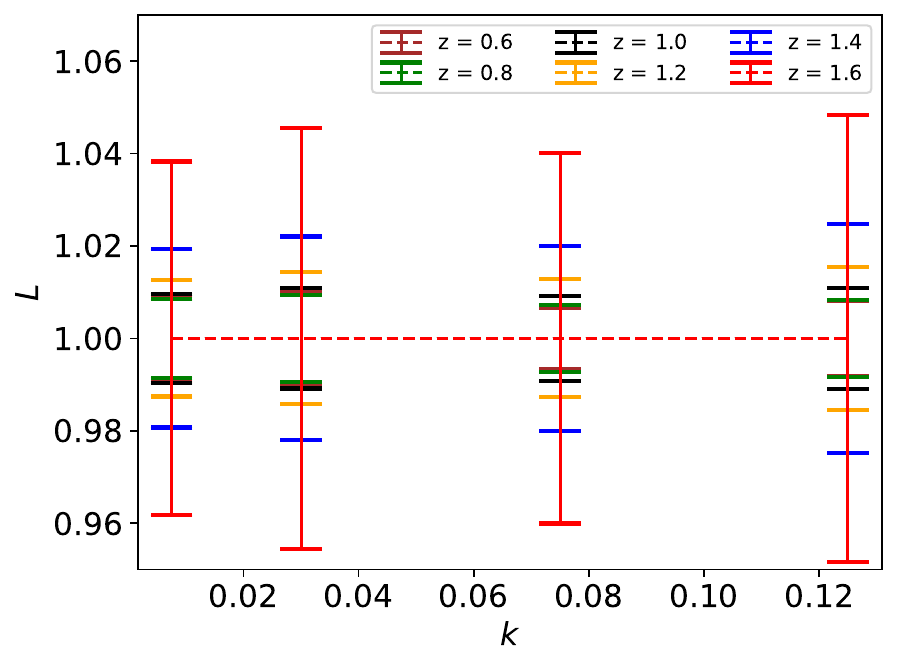}
\hfil
\includegraphics[width=0.35\textwidth]{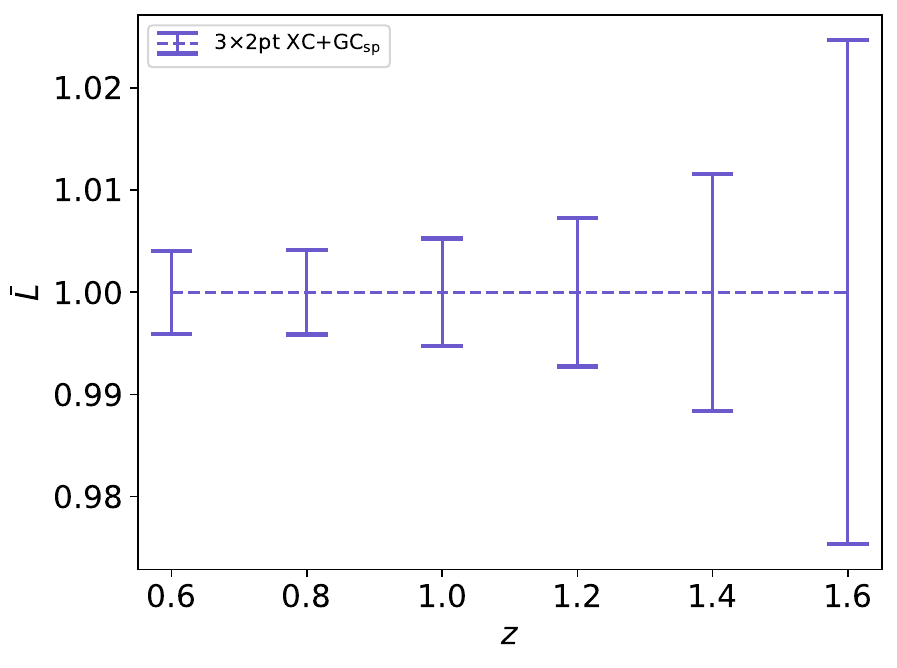}
\medskip
\includegraphics[width=0.35\textwidth]{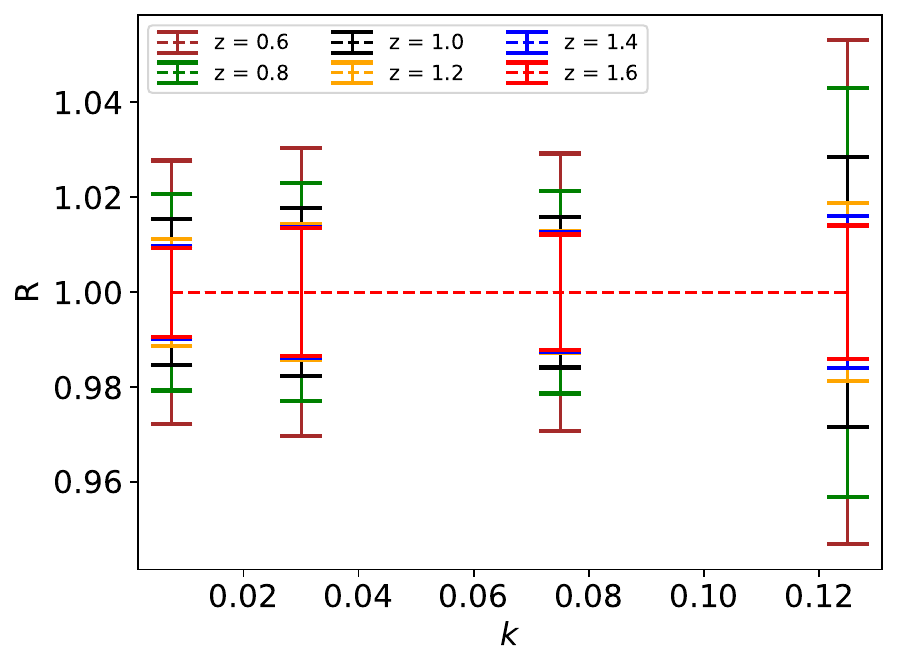}
\hfil
\includegraphics[width=0.35\textwidth]{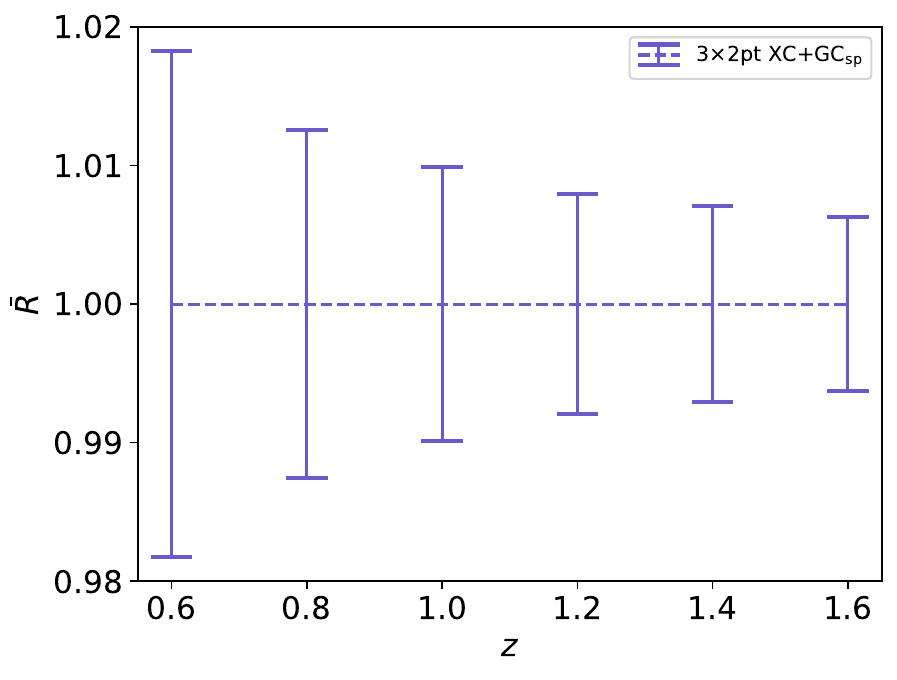}
\medskip
\includegraphics[width=0.35\textwidth]{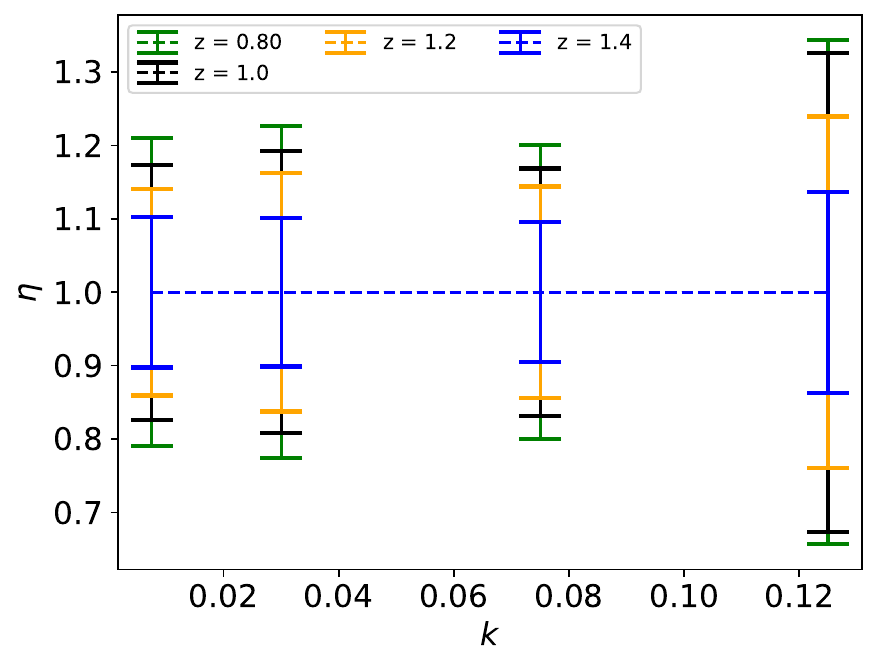}
\hfil
\includegraphics[width=0.35\textwidth]{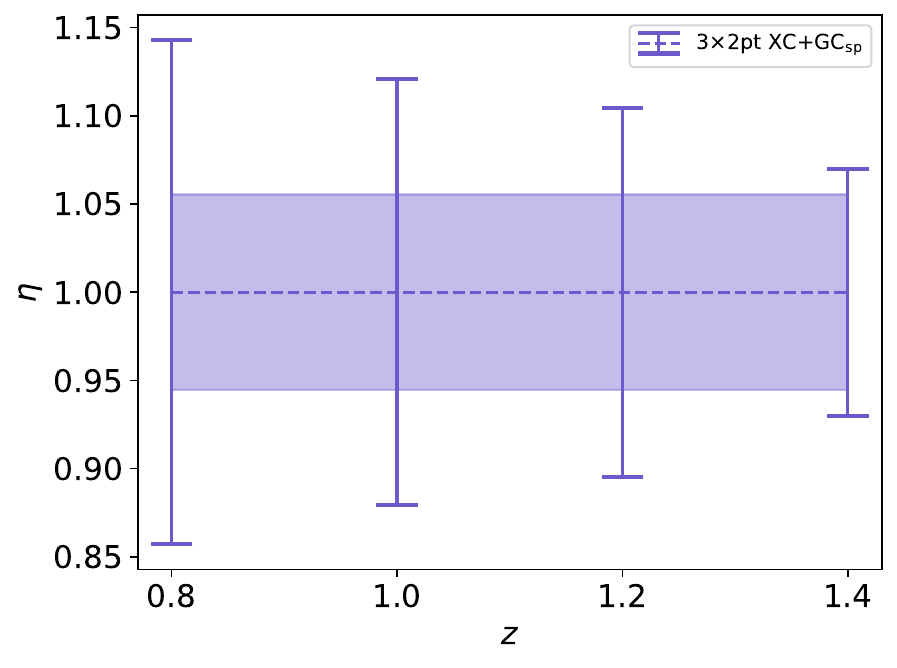}
\medskip
\includegraphics[width=0.35\textwidth]{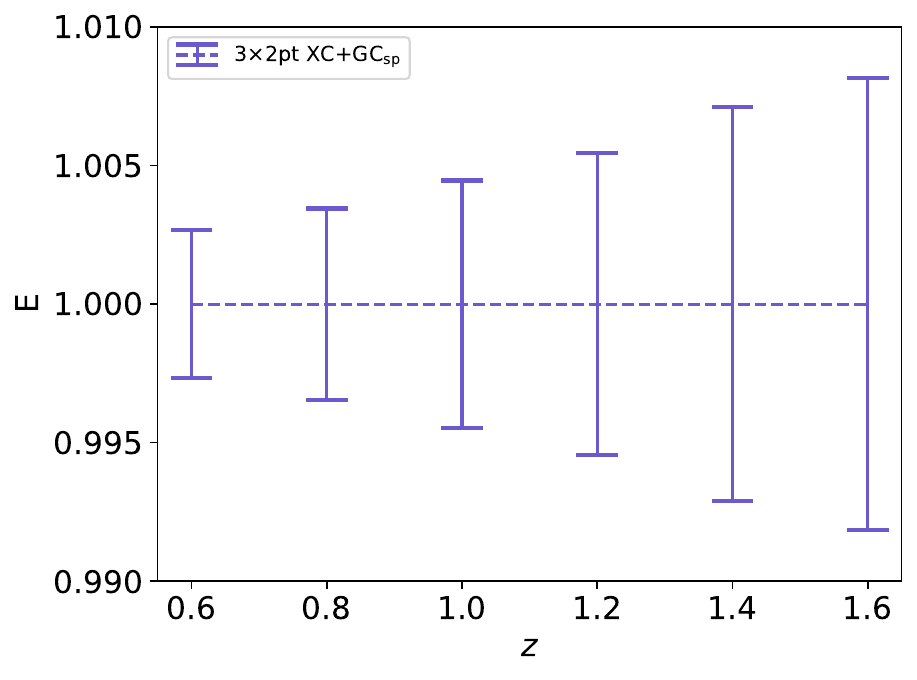}
\hfil
\caption{Relative error bars on $L$, $R$, $E$ and $\eta$ obtained from Fisher forecast using the photometric and spectroscopic observables, following the formalism described in Sect.~\ref{sec:photo_spectro_obs} and the settings described in Sect.~\ref{sec:fisher}. Left panel: results for all the  $z$ and $k$ bins. Right panel: assuming $z$-dependent binning only, while the last panel also shows in the shadow region errors when $\eta$ is assumed constant in all redshift bins. 
}
    \label{fig:baseline} 
\end{figure}

Following the method detailed above in Sect.~\ref{sec:photo_spectro_obs} and the settings described in Sect.~\ref{sec:fisher}, we present the predicted 1$\sigma$ errors for $\eta$ along with those for the intermediate parameters $E$, $R$ and $L$, after marginalizing over $A$ and the other nuisance parameters. We first show, in Fig.~\ref{fig:baseline}, what we consider as our baseline case, where all relative error bars are derived from Fisher forecasts using photometric and spectroscopic observables, including cross-correlations in the photo survey and accounting for the IA and the FoG effect. The left panel presents results for all the $z$ and $k$ bins, while the right panel shows error bars assuming $z$-dependent binning only for the parameters $\bar{R}=R/\delta_{\rm m,0}(k)$ and $\bar{L}=L/\delta_{\rm m,0}(k)$, as $\delta_{\rm m,0}(k)$ cannot be $k$ independent by definition. \footnote{To obtain $\bar{L}$ and $\bar{R}$, a fixed shape of $P(k)$ (such as the $\Lambda$CDM shape used here) must be assumed. Thus, these model-dependent quantities are evaluated only for better comparisons between the baseline and other tests. Note that this not the case for $\eta$ since $P_2$ and $P_3$ are independent from $\delta_{\rm m,0}$. However, we still loose, to a lesser degree though, in model independency when we consider the $z$ only dependent or the constant case for $\eta$ since $P_i$ are treated as space and time dependent in the first $z$ and $k$ binning case.}. Additionally, the bottom-right panel shows the scenario where $\eta$ is assumed constant in all redshift bins. All values from this baseline, along with other cases we discussed later, are summarized in table~\ref{tab:z-only}. 

Going through the different plots, we first observe relative errors for the $E(z_i)$ within or less than 1\%  with an increase of the errors for high redshifts. This is due to the fact that the $E(z_i)$ at lower $z$ take part more than the ones of the higher $z$ bins in the modeling of the projection of the lensing of the sources all the way up to the last observed bin. For the $R(z_i,k_i)$ parameter, we observe that errors are in the few percents range. Here we do not observe a decreasing trend with increasing wave numbers $k$, since we expect that $R$, essentially constrained by the spectroscopic measurements, will have in the corresponding cells in the Fisher Matrix lower values for low $k$, as we see from examining Eq.~\ref{eq:fmgc}. Indeed, we checked that this is the case if we calculate the marginalized errors only from the $R$ rows and columns in the Fisher matrix. Thus it remains that the marginalization when including the other parameters is what is mitigating this behaviour. While the errors on $\bar{R}$ after assuming $z$-only dependence improve by a factor of two on average with a decreasing trend with redshift. The latter is due to the fact that the IA effect that involves $R$ acts as an additional constraining factor with redshift to the one coming from the spectroscopic measurements using this parameter. The picture is not different for the $z, k$ binning for the $L$ parameters, where no significant trend was found as function of the wave-number, though still with values in the order of a few precent. However, in the $z$ only assumption, the trend goes with higher error bars with the redshift. This could be understood by the fact that $L$ is essentially constrained by the projected lensed spectra from the photometric measurements with a decreasing number of lenses when going up to higher redshift bins. The previous argument would explain the trend for $\eta$, whose bounds go from 10\% to 30\% in the $z, k$ binning with only weak variation with the wave-number, as was the case for $R$ and $L$. This is due mainly to the fact that our reconstruction method interpolates and smooth the $k$ dependence, but also as we shall see later, including galaxy angular power spectrum in our probes as well as the IA effect, both having all our parameters as ingredients, helps in reducing any privileged behaviour as function of $k$ for one of the probes vs another. When passing to the $z$ dependence, we also observe a decrease by a factor of 2, and a decreasing trend with redshift that is probably due to the fact that $P_2$ cancel $E$ errors in Eq.~\ref{eq:eta} leaving $R$ in $P_3$ as the ruler. Finally, when we project $\eta$ following the assumption of a constant value all over the redshift and the wave-number we observe a substantial gain, since we are now becoming more model dependent, reaching $\sim 5$ \% as seen in Table~\ref{tab:eta}. This is better than one order of magnitude from current constraints \citep{Planck:2018vyg, DES:2022ccp, Sakr:2023xnw} and in the same order as other model-dependent forecasts studies forecasting on $\eta$ from similar  surveys \citep{Martinelli:2021hir,Casas:2022vik}. Note that we checked, as a verification and robustness test, that other common reconstruction methods, e.g. linear instead of cubic interpolation, end up giving the same bounds on $\eta$ in the constant case. 

\begin{figure}[htbp]
    \centering
\includegraphics[width=0.35\textwidth]{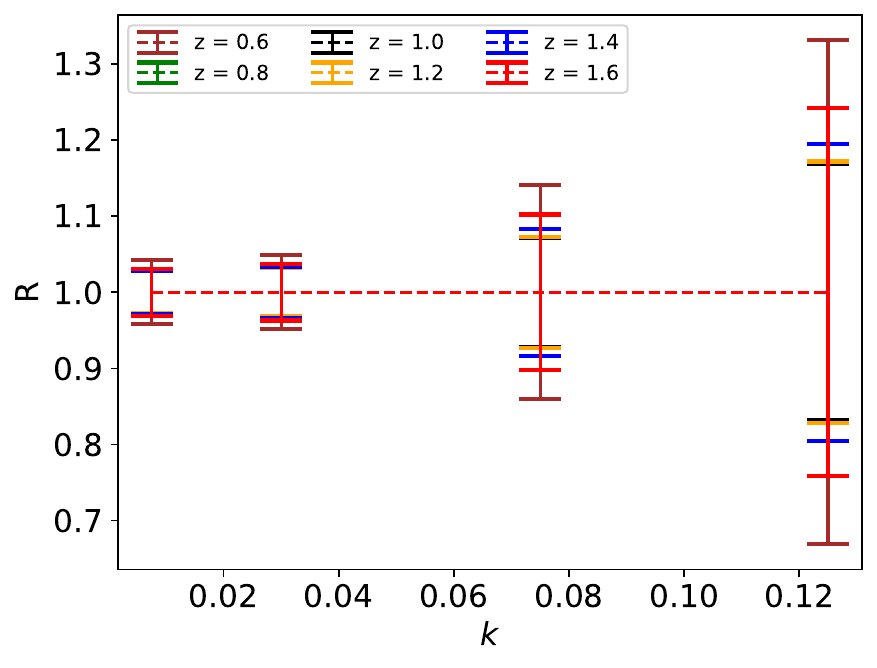}
\hfil
\includegraphics[width=0.35\textwidth]{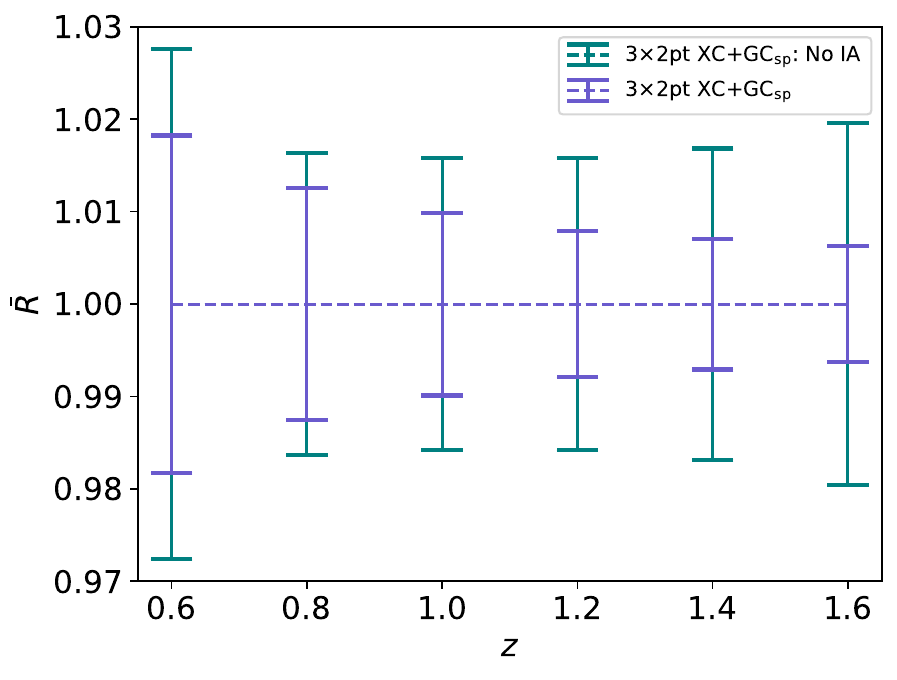}
\caption{Relative errors on $R$ following the same settings as in Fig.~\ref{fig:baseline}. Left panel: 1~$\sigma$ constraints on $R$ obtained without accounting for intrinsic alignment contamination in the lensing of galaxies. Right panel: comparison between $R(z)$ without intrinsic alignment and baseline configuration.}
\label{fig:err_combined_R_noIA}
\end{figure}

\begin{figure}[htbp]
    \centering
\includegraphics[width=0.3\textwidth]{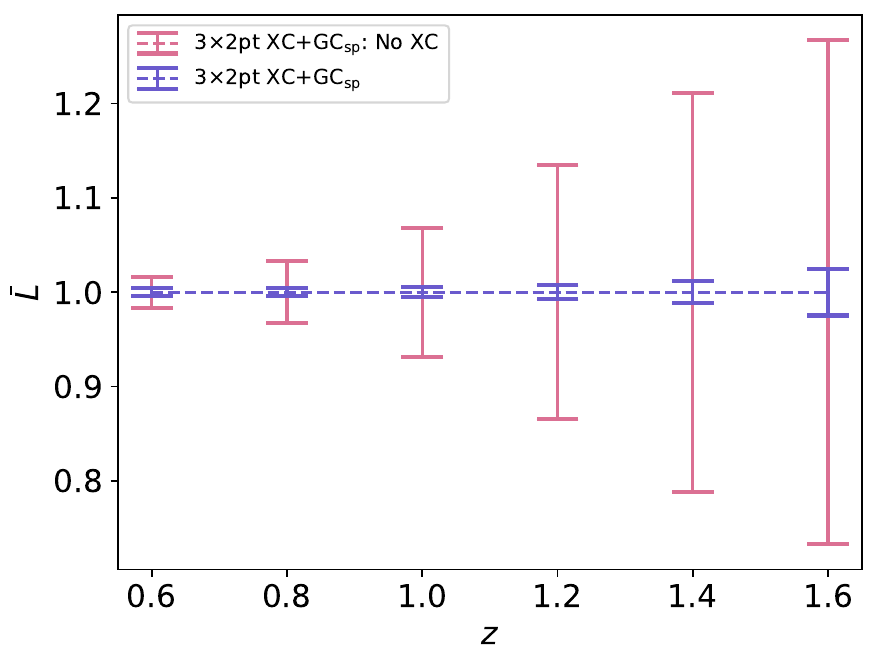}
\hfil
\includegraphics[width=0.3\textwidth]{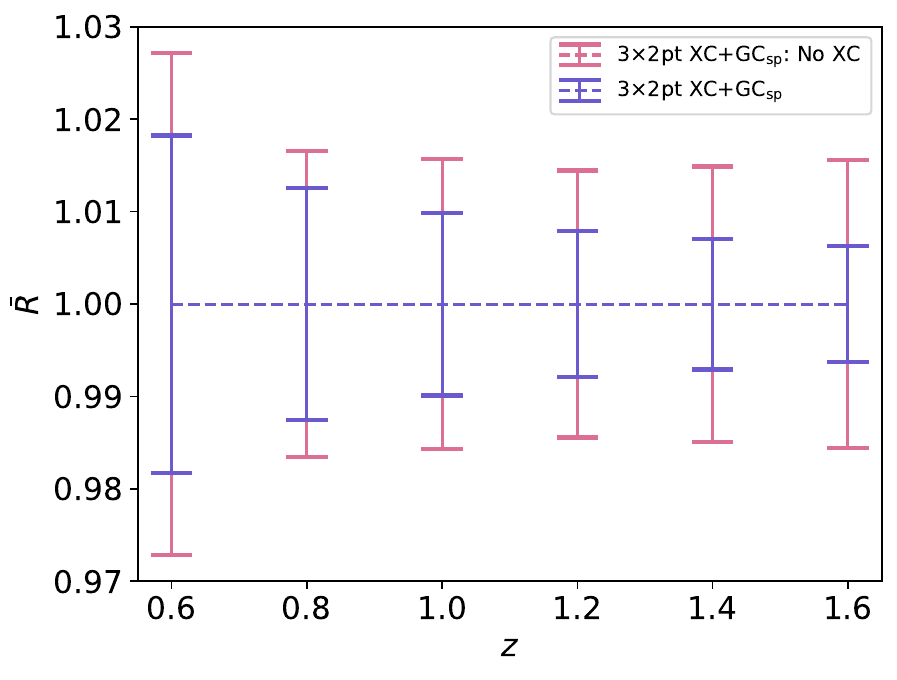}
\hfil
\includegraphics[width=0.3\textwidth]{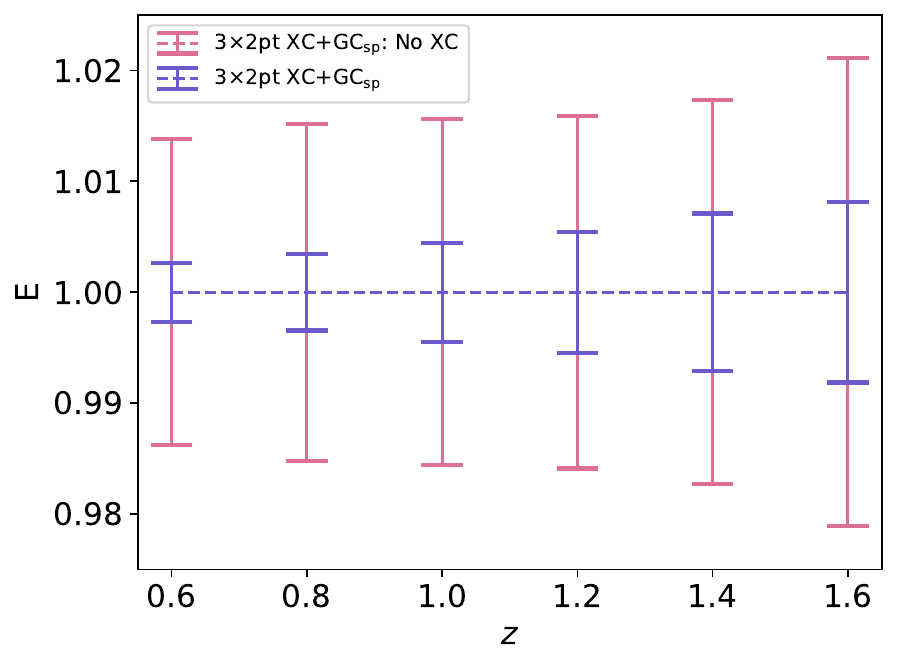}
\caption{ 1~$\sigma$ relative error on $L, R$ and $E$, asumming only $z$ dependence. We show the comparison between base line configuration and "No XC", i.e. without including photometrically detected galaxy-galaxy clustering, along with the galaxy lensing-lensing and their cross-correlated angular power spectrum.}
    \label{fig:err_combined_L_noXC}
\end{figure}

\begin{figure}[htbp]
    \centering
\includegraphics[width=0.35\textwidth]{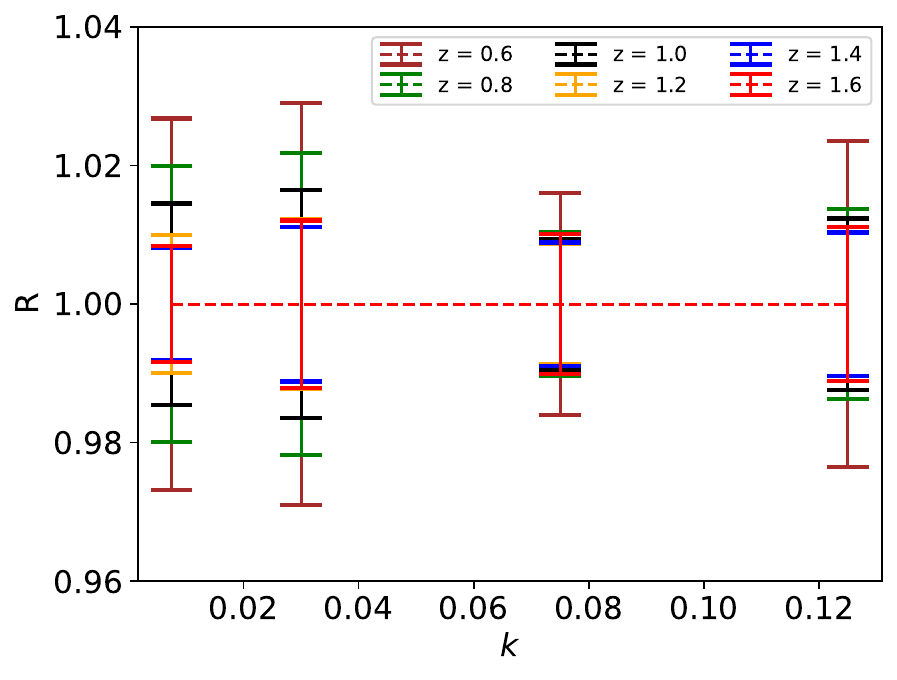}
\hfil
\includegraphics[width=0.35\textwidth]{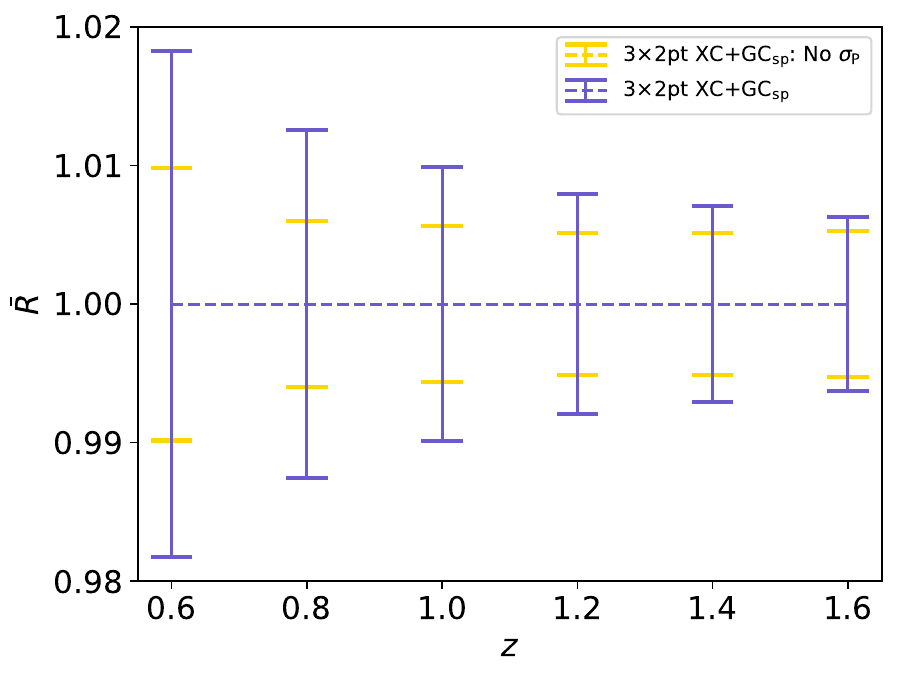}
\caption{1~$\sigma$ relative errors on $R$, assumming only $z$ dependence. In the right panel we show the comparison between baseline and "No $\sigma_{\rm p}$", i.e. without taking into account the Finger-of-God effect.}
    \label{fig:err_combined_R_nosigp}
\end{figure}

\begin{figure*}
\begin{center}
\includegraphics[width=2.2in,height=1.6in]{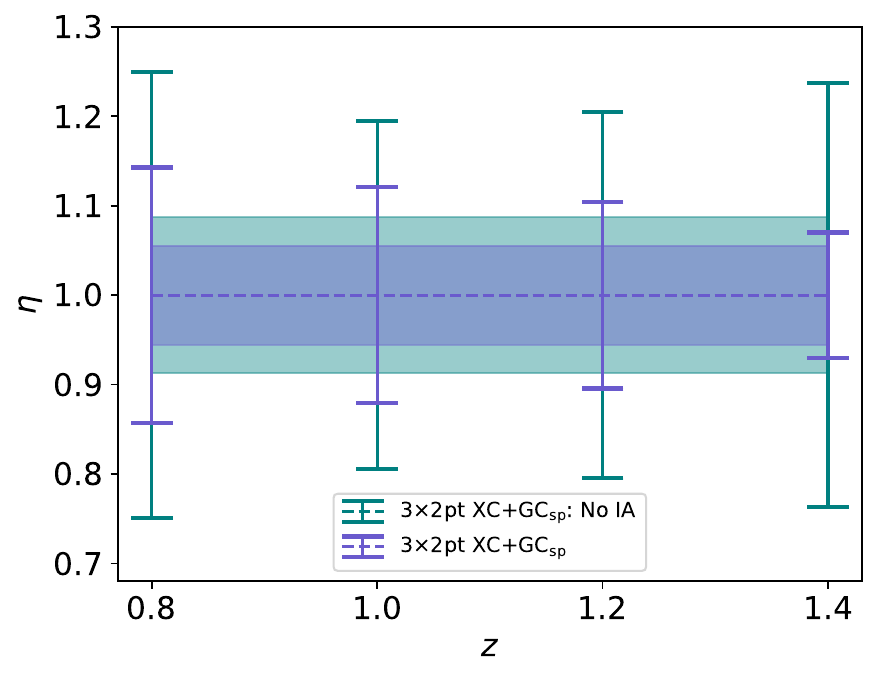}
\includegraphics[width=2.2in,height=1.6in]{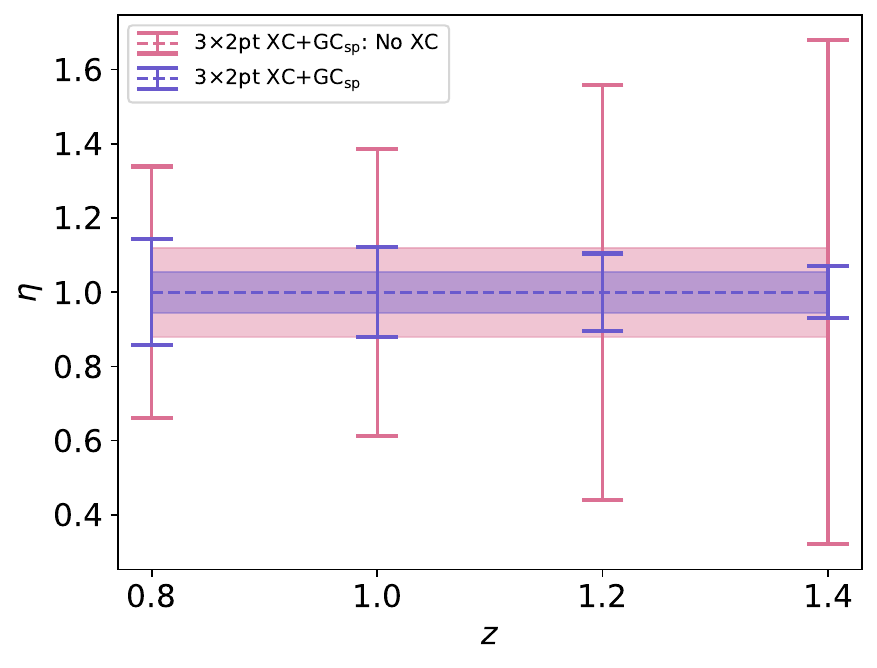}
\includegraphics[width=2.2in,height=1.6in]{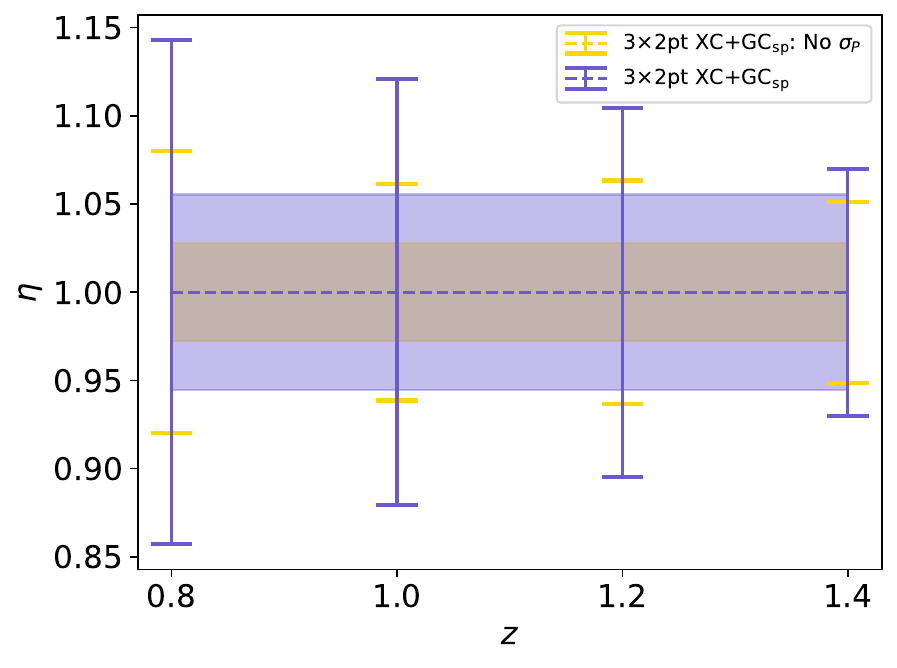}
\caption{Relative errors on $\eta$, following the same settings as in Fig.~\ref{fig:baseline} but assuming $z$ dependent binning only, while also showing the case where $\eta$ is assumed constant in all redshift bins. Left panel: comparison of the errors with and without accounting for the intrinsic alignment effect. Middle panel: comparison of the errors with and without including galaxy-galaxy clustering, in addition to their lensing and their cross-correlations. Right panel: comparison of the errors with and without taking into account the Finger-of-God effect.} \label{fig:eta_z_XC_IA_sigp}
\end{center}
\end{figure*}

To gain further insights, check our findings or try to isolate the contribution of improvement from modifications of this study with respect to previous ones, we now show other particular cases, such as the one where we do not account for the IA effect as in Fig.~\ref{fig:err_combined_R_noIA}, or one without including cross correlations from the photometric surveys (Fig.~\ref{fig:err_combined_L_noXC}) and only limiting to the galaxy galaxy lensing probes, or finally one where we neglect the nuisance from the FoG as in Fig.~\ref{fig:err_combined_R_nosigp}. We also group all the values in table~\ref{tab:z-only} next to the ones from our baseline. We show each time the parameters that were impacted the most from our choices with respect to the baseline. Therefore, we observe in Fig.~\ref{fig:err_combined_R_noIA} an increase in the error bars with respect to the baseline of almost one order of magnitude, due to the fact that $R$ is not any more constrained by the photo probes, following Eq.~\ref{eq:newmod_IA_Cell}, but only by the spectroscopic ones. A smaller difference in the order of 50\% with respect to the baseline is seen in the $z$ only assumption. This difference in $R$ translates in the final bounds on $\eta$ in Fig~\ref{fig:eta_z_XC_IA_sigp} where we find that we loose precision by the same order of magnitude for all redshifts as well in the $\eta$ constant model assumption case as we also see in table~\ref{tab:eta}. In the case where we do not include cross correlations and the galaxy-galaxy angular power spectrum in the photometric survey, we expect and see in table~\ref{tab:z-only} that the $R$ and $L$ parameters are impacted uniformly regardless of the wave-number, therefore we show in Fig.~\ref{fig:err_combined_L_noXC} the $z$ dependence for $R$, $L$ and $E$. We observe that the trend is conserved as noted and that $R$ is the least impacted since it gets its constrained from the spectroscopic probe and the IA which are both still present, while $L$ changes the most due to the fact that we are loosing in this case the power of the lensing effects from the galaxy - galaxy lensing correlations. This difference in $R$, $L$ and $E$ translates in the final bounds on $\eta$ in Fig~\ref{fig:eta_z_XC_IA_sigp} where we find that we loose precision by 50\% for all redshifts or in the case of the $\eta$ constant model assumption as we see in table~\ref{tab:eta}. Finally, neglecting the FoG effect as in Fig.~\ref{fig:err_combined_R_nosigp}, naturally impacts the $R$ parameter in its error bar values and show a trend in the $k,z$ dependence plot, since this nuisance only affects the spectroscopic probe as function of the wave-number following Eq.~\ref{gcpower}. This difference in $R$ translates in the final bounds on $\eta$ in Fig~\ref{fig:eta_z_XC_IA_sigp}, where we rather gain precision to more than 50\% in the $z$ dependent or the constant assumption shown in table~\ref{tab:eta}.  

\begin{table}
\centering
\begin{tblr}{|l|c c c c c c||c c c c c c|}
\toprule
   &&& \SetCell[c=2]{c} 3x2pt XC + GC$_{\rm sp}$ : No IA  & & & &\SetCell[c=6]{c} 3x2pt XC + GC$_{\rm sp}$ & \\ 
\midrule
$\sigma $ & $z = 0.6$ & $z = 0.8$ & $z=1.0$ & $z=1.2$ & $z=1.4$  & $z = 1.6$ & $z=0.6$ &  $z=0.8$ & $z=1.0$& $z=1.2$ & $z=1.4$ & $z=1.6$\\
\midrule
   $\bar{L}$ & 0.413\% & 0.415\% & 0.524\% & 0.713\% & 1.05\% & 1.89\% & 0.407\% & 0.414\% & 0.526\% & 0.727\% & 1.16\% & 2.47\%\\
   $\bar{R}$ & 2.76\% & 1.63\% & 1.58\% & 1.58\% & 1.68\% & 1.96\% & 1.82\% & 1.26\% & 0.99\% & 0.792\% & 0.708\% & 0.627\%\\ 
   $E$ & 0.272\% & 0.357\% & 0.482\% & 0.607\% & 0.894\% & 1.04\% & 0.266\% & 0.345\% & 0.446\% & 0.546\% & 0.71\% & 0.815\%\\ 
   $\eta$ & - & 24.9\% & 19.4\% & 20.5\%, & 23.7\% & - & -  & 14.3\% & 12.1\% & 10.5\% & 7.00\% & - \\ 
\hline 
\hline
  &&& \SetCell[c=2]{c} 3x2pt XC + GC$_{\rm sp}$ : No XC  & & & &\SetCell[c=6]{c} 3x2pt XC + GC$_{\rm sp}$ : No $\sigma_{\rm p}$ & \\ 
\midrule
$\sigma$ & $z = 0.6$ & $z = 0.8$ & $z=1.0$ & $z=1.2$ & $z=1.4$  & $z = 1.6$ & $z=0.6$ &  $z=0.8$ & $z=1.0$& $z=1.2$ & $z=1.4$ & $z=1.6$\\
\midrule
   $\bar{L}$ & 1.64\% & 3.29\% & 6.85\% & 13.5\% & 21.1\% & 26.7\% & 0.402\% & 0.412\% & 0.523\% & 0.719\% & 1.13\% & 2.39\% \\
   $\bar{R}$ & 2.72\% & 1.65\% & 1.57\% & 1.44\% & 1.49\% & 1.56\% & 0.984\% & 0.56\% & 0.565\% & 0.513\% & 0.511\% & 0.525\% \\ 
   $E$ &1.38\% & 1.52\% & 1.56\% & 1.59\% & 1.73\% & 2.11\% & 0.264\% & 0.341\% & 0.441\% & 0.535\% & 0.687\% & 0.788\%\\ 
   $\eta$ & - & 33.9\% & 38.7\% & 55.9\% & 67.9\% & - & - & 7.98\% & 6.14\% & 6.33\% & 5.13\% & - \\    
\midrule
\end{tblr}
\caption{1$\sigma$ relative errors for $L, R, E$ and $\eta$, obtained from model-independent measurements, assuming these quantities depend only on $z$, where "No IA" refers to the case without accounting for the intrinsic alignment contamination in the lensing of galaxies, "No $\sigma_{\rm p}$" refers to the case without accounting for the Finger-of-God effect and "No XC" refers to when we limit our forecast in its photometric probes to the galaxy lensing-lensing angular power spectrum.}
\label{tab:z-only}
\end{table}

\begin{table}
\centering
\begin{tblr}{|l|c c c c||c c c c|}
\toprule
  && \SetCell[c=2]{c} 3x2pt XC + GC$_{\rm sp}$ : No IA  & & &\SetCell[c=4]{c} 3x2pt XC + GC$_{\rm sp}$  & \\ 
\midrule
$\sigma_\eta$ & $z = 0.8$ & $z = 1.0$ & $z=1.2$ & $z=1.4$ & $z=0.8$ 
& $z = 1.0$ & $z=1.2$ 
& $z=1.4$ \\ 
\midrule
   $k=0.0075$ & 0.351 & 0.299 & 0.310 &  0.342 & 0.210 & 0.174 & 0.141 & 0.103\\
   $k=0.03$ & 0.402 & 0.336 & 0.350 & 0.383 & 0.226 &  0.192  & 0.162 & 0.101\\ 
   $k=0.075$ & 1.107 & 0.751 & 0.846 & 1.027 & 0.200 & 0.169 & 0.144 & 0.095\\ 
   $k=0.125$ & 2.613 & 1.790 & 2.041 & 2.507 & 0.343  & 0.326   & 0.239 & 0.137\\ 
\midrule
   $\eta_{\rm const}$  && \SetCell[c=2]{c} 0.088  & & &\SetCell[c=4]{c} 0.055 & \\
\hline 
\hline
 && \SetCell[c=2]{c} 3x2pt XC + GC$_{\rm sp}$ : No XC  &&& \SetCell[c=4]{c} 3x2pt XC + GC$_{\rm sp}$ : No $\sigma_{\rm p}$ &  \\ 
\midrule
$\sigma_\eta$ & $z = 0.8$ & $z = 1.0$ & $z=1.2$ & $z=1.4$ & $z=0.8$  & $z = 1.0$ & $z=1.2$ &  $z=1.4$ \\ 
\midrule
   $k=0.0075$ & 0.494 &  0.584 & 0.793 & 0.895 & 0.207  & 0.166 & 0.131 & 0.098\\
   $k=0.03$ & 0.547 &  0.697  & 1.009 & 1.248 & 0.222 & 0.181  & 0.144  & 0.093\\ 
   $k=0.075$ & 0.518 &  0.673 & 0.950 &  1.200 & 0.126 &  0.093 & 0.090 &  0.076\\ 
   $k=0.125$ & 0.873 &  1.010  & 1.164 & 1.165 & 0.184 &  0.125   & 0.115 & 0.091\\    
\midrule
   $\eta_{\rm const}$  && \SetCell[c=2]{c} 0.12  & & &\SetCell[c=4]{c} 0.028 & \\
\midrule
\end{tblr}
\caption{1$\sigma$ percentage relative errors values on $\eta$, obtained from model-independent measurements at various $z$ and $k$ bins along with the relative error when considering $\eta$ as constant along the whole redshift and wave-number range, where "No IA" refers to the case without accounting for the intrinsic alignment contamination in the lensing of galaxies, "No $\sigma_{\rm p}$" refers to the case without accounting for the Finger-of-God effect and "No XC" refers to when we limit our forecast in its photometric probes to the galaxy lensing-lensing angular power spectrum.}
\label{tab:eta}
\end{table}

\newpage
\section{Conclusions}\label{sec:conclusion}

In this paper, we present a model-independent forecast of constraints on the anisotropic stress, $\eta$, for future large-scale surveys that combine spectroscopic galaxy clustering and weak lensing measurements. We also employ photometric observation of projected lensing and galaxy clustering correlations, along with their cross-signals, to estimate $\eta$ from three directly observable functions of scale and redshift that depend on the cosmic expansion rate $E$, on the linear growth rate $R$, and on the lensing correlation $L$ in a way that is independent of assumptions about background cosmology, galaxy bias, initial conditions, and matter abundance. For the photometric sample, we choose specifications for a Euclid-like survey, while for the spectroscopic survey, we join a DESI-like survey at low redshift to a Euclid-like one at higher redshift. We consider three scenarios: $\eta$ and its forming components as a free function of both redshift and scale, $\eta$ with redshift dependence only, and a constant $\eta$ along all bins. In our baseline case, i.e. when including galaxy clustering and cross-correlations with galaxy galaxy lensing, and accounting for IA and FoG, we found in the $z$ dependence case that $L$ and $R$ error bars are below 2\% for all bins, while showing no preference for a specific wave number in the $z, k$ binning. We also found that $E$ could be constrained to less than 1\%. Finally, our targeted parameter $\eta$ had relative error range between 10 and 20\% in the $z$ dependent case, to reach $\sim$ 5\% when considered constant for all $z$ and $k$ bins. The latter degrades by almost 50\% when IA is not included with the main impact coming from the $R$ parameter that is now only constrained by the spectroscopic observables. A similar gain is obtained on $\eta$ relative errors without the FoG nuisance, with impact from the same parameter $R$ since this nuisance is relative to the spectroscopic observed power spectrum. Finally, not including XC in our probes impacts all our intermediate parameters $L$, $R$ and $E$, albeit much more strongly on the lensing one, which result into a degradation in the order of a factor of 2 on the relative errors on $\eta$. We also investigated, within our baseline configuration, different cases where we do not include $C_{\ell s}$ from redshifts below the range of our model binning, or those from $z$ higher than the limit of our last bin. We found that the strongest impact comes from the angular correlations of the high $z$ galaxies, especially on the $L$ parameter, resulting in an increase of a factor of 2 on the error on $\eta$. We conclude that, despite the strong capabilities of the next generation surveys, $\eta$ in the most model independent considerations, i.e. in the $z, k$ binning scheme, will only be constrained on average around 15\%, still leaving room for various alternative gravity and dark energy models. We also emphasize on the power of the XC in helping to improve the constraints and the importance of accounting for the nuisance effects for more accurate results. Finally, we note that our study was conducted with still being limited to linear scales and future works should address introducing non linear scales within our model independent approaches to harvest more the power of the upcoming Stage-IV surveys.

\section*{Acknowledgements}

The authors would like to thank Luca Amendola for his useful comments and discussions. ZS acknowledges support from the DFG project 456622116. ZZ acknowledges
support from DFG Germany’s Excellence Strategy EXC 2181/1
- 390900948 (the Heidelberg STRUCTURES Excellence
Cluster).

%%%%%%%%%%%%%%%%%%%%%%%%%%%%%%%%%%%%%%%%%%%%%%%%%%
% \section*{Data Availability}

% No data was needed or used in this forecast study.



\bibliographystyle{mnras}
\bibliography{example} % if your bibtex file is called example.bib

\begin{thebibliography}{}
\makeatletter
\relax
\def\mn@urlcharsother{\let\do\@makeother \do\$\do\&\do\#\do\^\do\_\do\%\do\~}
\def\mn@doi{\begingroup\mn@urlcharsother \@ifnextchar [ {\mn@doi@}
  {\mn@doi@[]}}
\def\mn@doi@[#1]#2{\def\@tempa{#1}\ifx\@tempa\@empty \href
  {http://dx.doi.org/#2} {doi:#2}\else \href {http://dx.doi.org/#2} {#1}\fi
  \endgroup}
\def\mn@eprint#1#2{\mn@eprint@#1:#2::\@nil}
\def\mn@eprint@arXiv#1{\href {http://arxiv.org/abs/#1} {{\tt arXiv:#1}}}
\def\mn@eprint@dblp#1{\href {http://dblp.uni-trier.de/rec/bibtex/#1.xml}
  {dblp:#1}}
\def\mn@eprint@#1:#2:#3:#4\@nil{\def\@tempa {#1}\def\@tempb {#2}\def\@tempc
  {#3}\ifx \@tempc \@empty \let \@tempc \@tempb \let \@tempb \@tempa \fi \ifx
  \@tempb \@empty \def\@tempb {arXiv}\fi \@ifundefined
  {mn@eprint@\@tempb}{\@tempb:\@tempc}{\expandafter \expandafter \csname
  mn@eprint@\@tempb\endcsname \expandafter{\@tempc}}}

\bibitem[\protect\citeauthoryear{Abbott et~al.}{Abbott
  et~al.}{2022}]{DES:2021wwk}
Abbott T. M.~C.,  et~al., 2022, \mn@doi [Phys. Rev. D]
  {10.1103/PhysRevD.105.023520}, 105, 023520

\bibitem[\protect\citeauthoryear{Abbott et~al.}{Abbott
  et~al.}{2023}]{DES:2022ccp}
Abbott T. M.~C.,  et~al., 2023, \mn@doi [Phys. Rev. D]
  {10.1103/PhysRevD.107.083504}, 107, 083504

\bibitem[\protect\citeauthoryear{Adame et~al.}{Adame
  et~al.}{2024}]{DESI:2024mwx}
Adame A.~G.,  et~al., 2024

\bibitem[\protect\citeauthoryear{Aghamousa et~al.}{Aghamousa
  et~al.}{2016}]{DESI:2016fyo}
Aghamousa A.,  et~al., 2016

\bibitem[\protect\citeauthoryear{Aghanim et~al.}{Aghanim
  et~al.}{2020}]{Planck:2018vyg}
Aghanim N.,  et~al., 2020, \mn@doi [Astron. Astrophys.]
  {10.1051/0004-6361/201833910}, 641, A6

\bibitem[\protect\citeauthoryear{Albuquerque et~al.}{Albuquerque
  et~al.}{2024}]{Albuquerque:2024}
Albuquerque et~al., 2024, in prep

\bibitem[\protect\citeauthoryear{Amendola, Kunz, Motta, Saltas  \&
  Sawicki}{Amendola et~al.}{2013}]{Amendola:2012ky}
Amendola L.,  Kunz M.,  Motta M.,  Saltas I.~D.,   Sawicki I.,  2013, \mn@doi
  [Phys. Rev. D] {10.1103/PhysRevD.87.023501}, 87, 023501

\bibitem[\protect\citeauthoryear{Amendola, Fogli, Guarnizo, Kunz  \&
  Vollmer}{Amendola et~al.}{2014}]{Amendola:2013qna}
Amendola L.,  Fogli S.,  Guarnizo A.,  Kunz M.,   Vollmer A.,  2014, \mn@doi
  [Phys.Rev.] {10.1103/PhysRevD.89.063538}, D89, 063538

\bibitem[\protect\citeauthoryear{Amendola, Bettoni, Pinho  \& Casas}{Amendola
  et~al.}{2020}]{Amendola:2019laa}
Amendola L.,  Bettoni D.,  Pinho A.~M.,   Casas S.,  2020, \mn@doi [Universe]
  {10.3390/universe6020020}, 6, 20

\bibitem[\protect\citeauthoryear{Amendola, Pietroni  \& Quartin}{Amendola
  et~al.}{2022}]{Amendola:2022vte}
Amendola L.,  Pietroni M.,   Quartin M.,  2022, \mn@doi [JCAP]
  {10.1088/1475-7516/2022/11/023}, 11, 023

\bibitem[\protect\citeauthoryear{Bacon et~al.}{Bacon
  et~al.}{2020}]{SKA:2018ckk}
Bacon D.~J.,  et~al., 2020, \mn@doi [Publ. Astron. Soc. Austral.]
  {10.1017/pasa.2019.51}, 37, e007

\bibitem[\protect\citeauthoryear{{Blanchard}, {Camera}, {Carbone}
  et~al.}{{Blanchard} et~al.}{2020}]{Euclid:2019clj}
{Blanchard} A.,  {Camera} S.,  {Carbone} C.,   et~al., 2020, \mn@doi [A\&A]
  {10.1051/0004-6361/202038071}, 642, A191

\bibitem[\protect\citeauthoryear{Casas, Carucci, Pettorino, Camera  \&
  Martinelli}{Casas et~al.}{2023}]{Casas:2022vik}
Casas S.,  Carucci I.~P.,  Pettorino V.,  Camera S.,   Martinelli M.,  2023,
  \mn@doi [Phys. Dark Univ.] {10.1016/j.dark.2022.101151}, 39, 101151

\bibitem[\protect\citeauthoryear{Catelan, Kamionkowski  \& Blandford}{Catelan
  et~al.}{2001}]{Catelan:2000vm}
Catelan P.,  Kamionkowski M.,   Blandford R.~D.,  2001, \mn@doi [Mon. Not. Roy.
  Astron. Soc.] {10.1046/j.1365-8711.2001.04105.x}, 320, L7

\bibitem[\protect\citeauthoryear{Clerkin, Kirk, Lahav, Abdalla  \&
  Gaztanaga}{Clerkin et~al.}{2015}]{Clerkin:2014pea}
Clerkin L.,  Kirk D.,  Lahav O.,  Abdalla F.~B.,   Gaztanaga E.,  2015, \mn@doi
  [Mon. Not. Roy. Astron. Soc.] {10.1093/mnras/stu2754}, 448, 1389

\bibitem[\protect\citeauthoryear{Hahn et~al.}{Hahn et~al.}{2023}]{Hahn:2022dnf}
Hahn C.,  et~al., 2023, \mn@doi [Astron. J.] {10.3847/1538-3881/accff8}, 165,
  253

\bibitem[\protect\citeauthoryear{Hirata \& Seljak}{Hirata \&
  Seljak}{2004}]{Hirata:2004gc}
Hirata C.~M.,  Seljak U.,  2004, \mn@doi [Phys. Rev. D]
  {10.1103/PhysRevD.82.049901}, 70, 063526

\bibitem[\protect\citeauthoryear{Li \& Xia}{Li \& Xia}{2025}]{Li:2025mib}
Li S.,  Xia J.-Q.,  2025

\bibitem[\protect\citeauthoryear{Martinelli \& Casas}{Martinelli \&
  Casas}{2021}]{Martinelli:2021hir}
Martinelli M.,  Casas S.,  2021, \mn@doi [Universe] {10.3390/universe7120506},
  7, 506

\bibitem[\protect\citeauthoryear{Mellier et~al.}{Mellier
  et~al.}{2024}]{Euclid:2024yrr}
Mellier Y.,  et~al., 2024

\bibitem[\protect\citeauthoryear{Pinho, Casas  \& Amendola}{Pinho
  et~al.}{2018}]{Pinho:2018unz}
Pinho A.~M.,  Casas S.,   Amendola L.,  2018, \mn@doi [JCAP]
  {10.1088/1475-7516/2018/11/027}, 11, 027

\bibitem[\protect\citeauthoryear{Raveri et~al.,}{Raveri
  et~al.}{2023}]{Raveri:2021dbu}
Raveri M.,  et~al., 2023, \mn@doi [JCAP] {10.1088/1475-7516/2023/02/061}, 02,
  061

\bibitem[\protect\citeauthoryear{Sakr}{Sakr}{2023}]{Sakr:2023xnw}
Sakr Z.,  2023, \mn@doi [JCAP] {10.1088/1475-7516/2023/08/080}, 08, 080

\bibitem[\protect\citeauthoryear{Sakr \& Martinelli}{Sakr \&
  Martinelli}{2022}]{Sakr:2021ylx}
Sakr Z.,  Martinelli M.,  2022, \mn@doi [JCAP] {10.1088/1475-7516/2022/05/030},
  05, 030

\bibitem[\protect\citeauthoryear{Troxel \& Ishak}{Troxel \&
  Ishak}{2014}]{Troxel:2014dba}
Troxel M.~A.,  Ishak M.,  2014, \mn@doi [Phys. Rept.]
  {10.1016/j.physrep.2014.11.001}, 558, 1

\bibitem[\protect\citeauthoryear{Tutusaus, Bonvin  \& Grimm}{Tutusaus
  et~al.}{2024}]{Tutusaus:2023aux}
Tutusaus I.,  Bonvin C.,   Grimm N.,  2024, \mn@doi [Nature Commun.]
  {10.1038/s41467-024-53363-6}, 15, 9295

\bibitem[\protect\citeauthoryear{Wang, Chuang  \& Hirata}{Wang
  et~al.}{2013}]{Wang:2012bx}
Wang Y.,  Chuang C.-H.,   Hirata C.~M.,  2013, \mn@doi [Mon. Not. Roy. Astron.
  Soc.] {10.1093/mnras/stt068}, 430, 2446

\bibitem[\protect\citeauthoryear{Zheng, Sakr  \& Amendola}{Zheng
  et~al.}{2024}]{Zheng:2023yco}
Zheng Z.,  Sakr Z.,   Amendola L.,  2024, \mn@doi [Phys. Lett. B]
  {10.1016/j.physletb.2024.138647}, 853, 138647

\makeatother
\end{thebibliography}

%%%%%%%%%%%%%%%%%%%%%%%%%%%%%%%%%%%%%%%%%%%%%%%%%%

%%%%%%%%%%%%%%%%% APPENDICES %%%%%%%%%%%%%%%%%%%%%

\appendix

\section{Robustness and cut in redshift bins}\label{sec:Robustness}

\begin{figure}[htbp]
\centering
\includegraphics[width=0.30\textwidth]{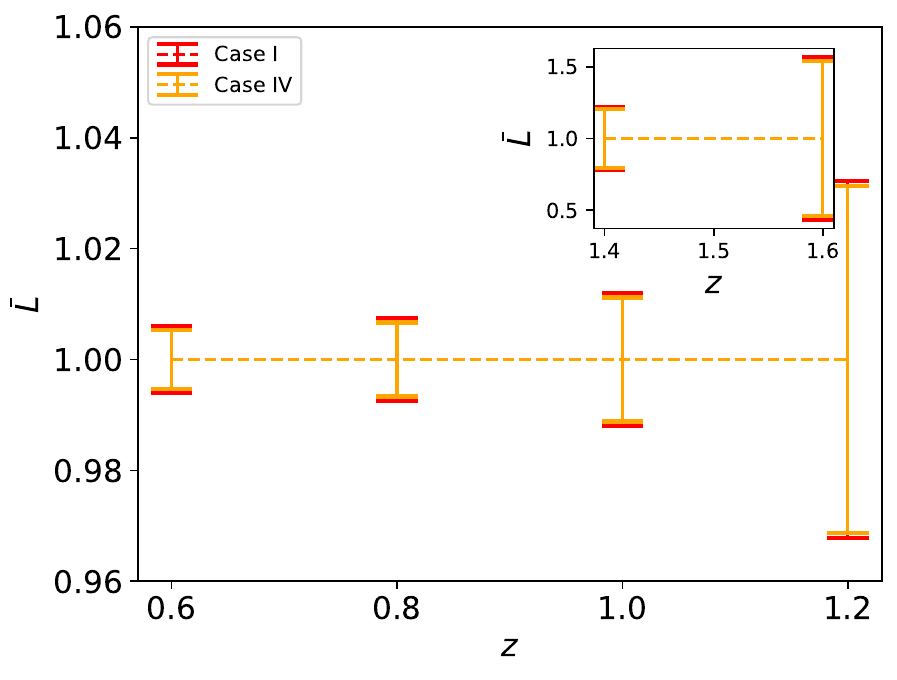}
\hfil
\includegraphics[width=0.30\textwidth]{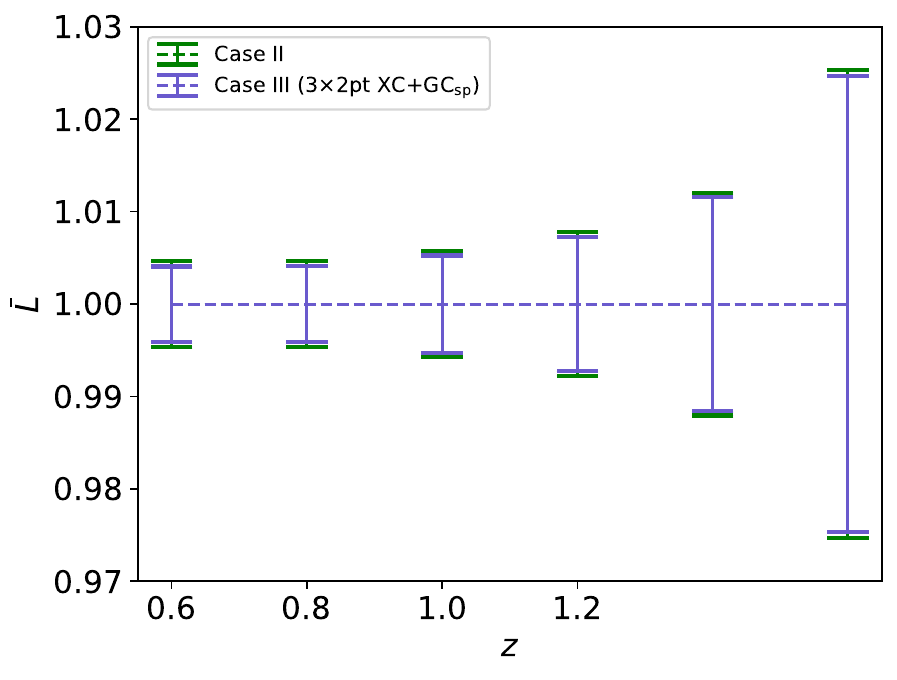}
\medskip \\
\includegraphics[width=0.30\textwidth]{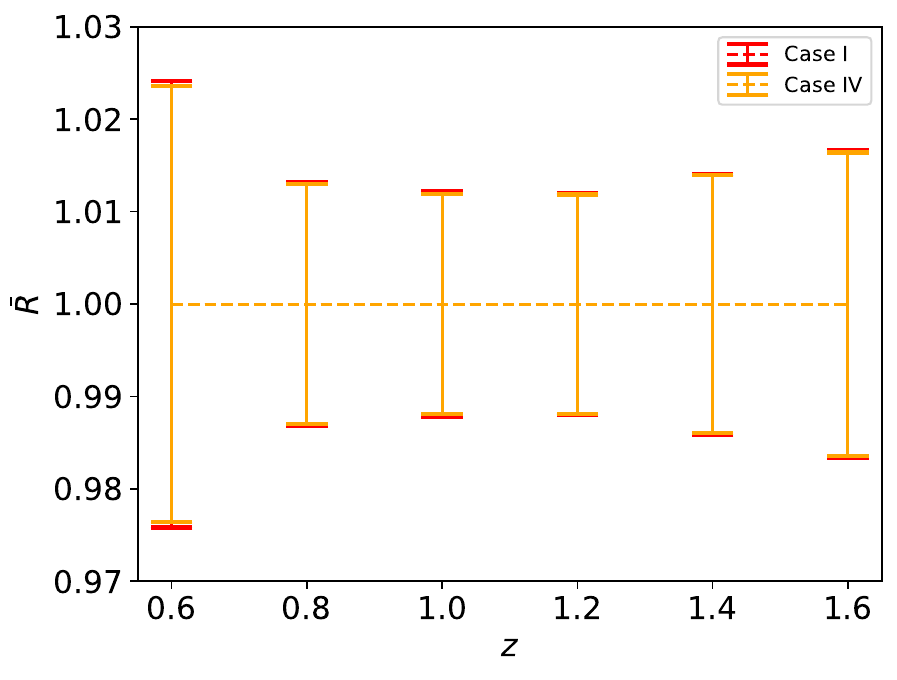}
\hfil
\includegraphics[width=0.30\textwidth]{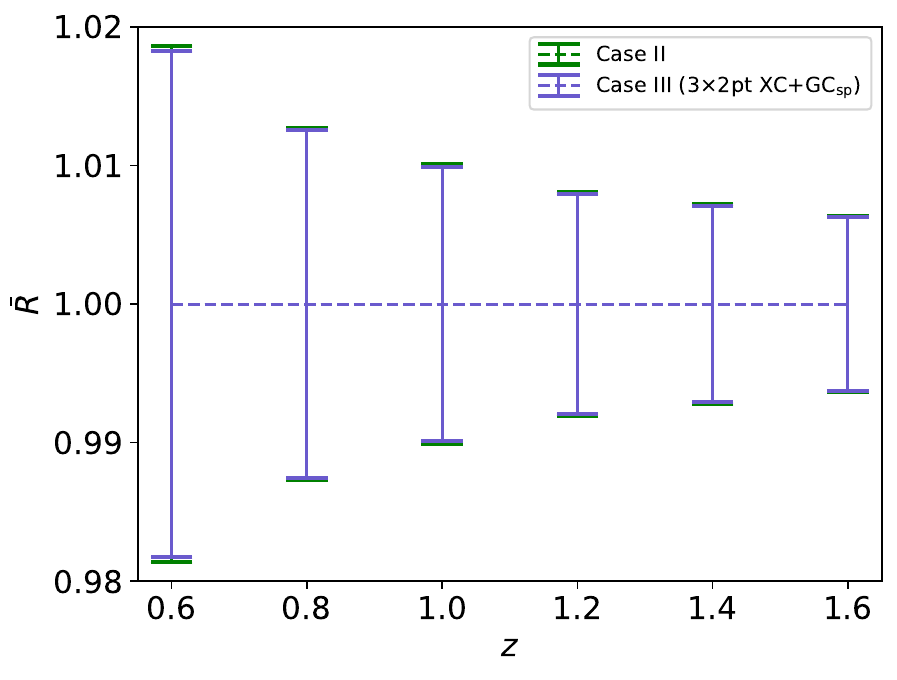}
\medskip \\
\includegraphics[width=0.30\textwidth]{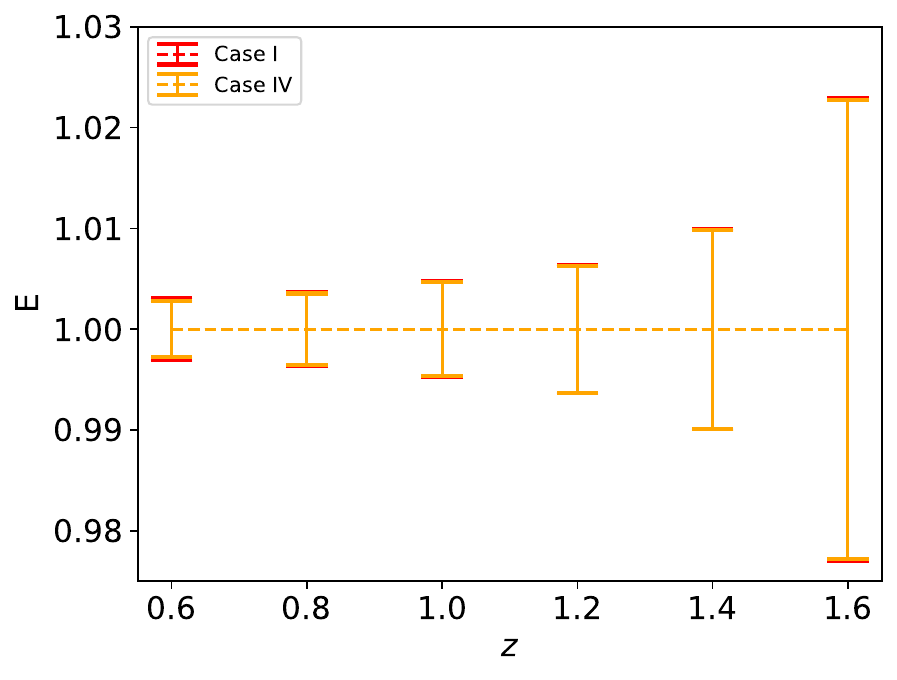}
\hfil
\includegraphics[width=0.30\textwidth]{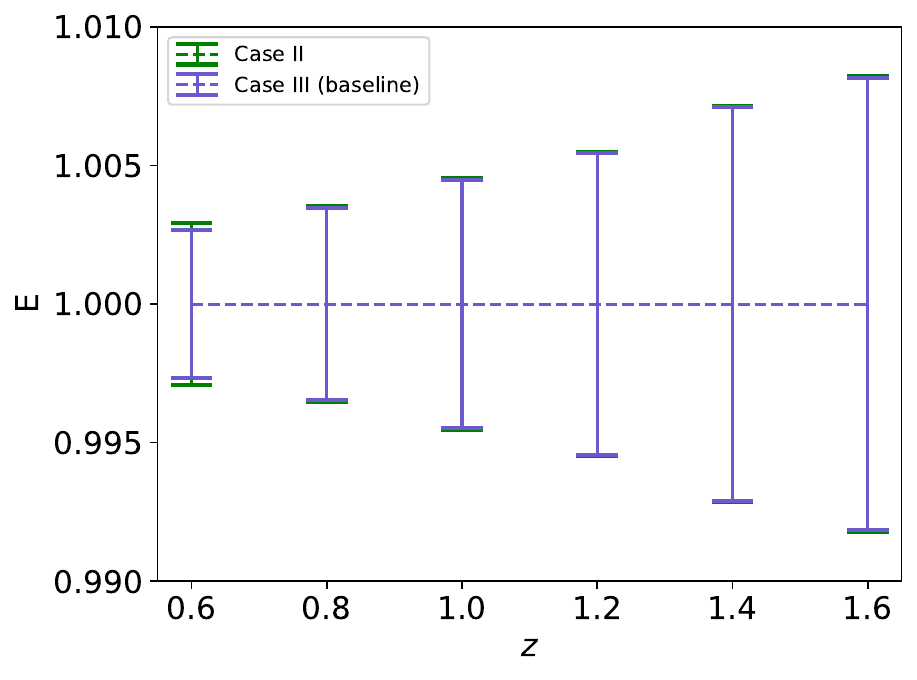}
\medskip \\
\includegraphics[width=0.30\textwidth]{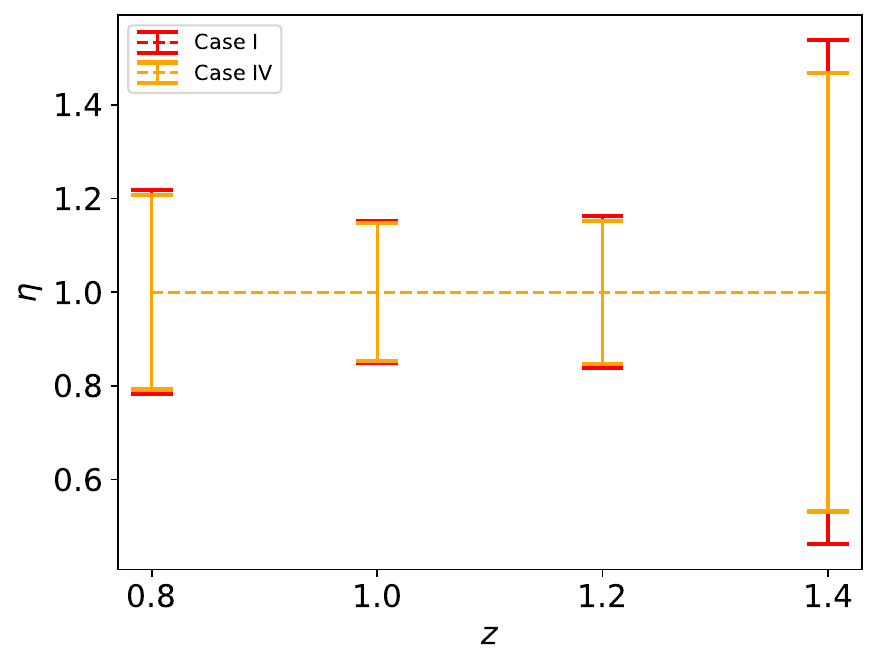}
\hfil
\includegraphics[width=0.30\textwidth]{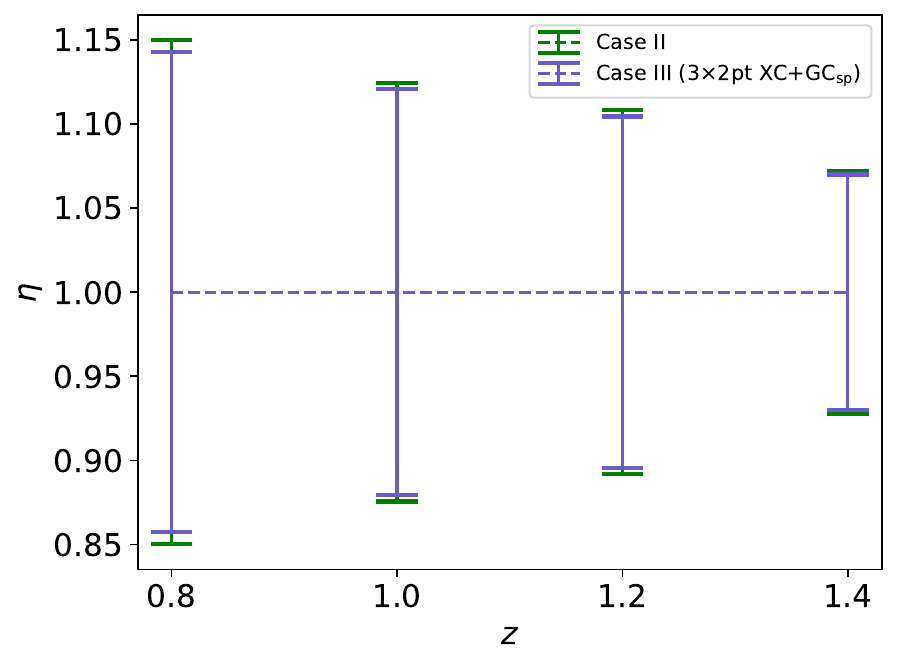}
\medskip
\caption{1~$\sigma$ relative errors on $L, R, E, \eta$, assuming only $z$ dependence. The four cases are summarized in Table \ref{tab:fourcases}. For $L$ in Case I and Case IV, the results in the last two bins shown in the embedded small plot, as they are significantly larger compared to the values in the other bins.}.
    \label{fig:cases} 
\end{figure}

The forecast conducted above was done considering all the lensed galaxies in the observed redshift range for the photometric survey. However, our redshift binning was limited so that to match the restricted binning range used for the spectroscopic survey. To get the ingredients or calculate the observables outside this range, we had to use of course an interpolation scheme. That could imply less model independency since we did not use all the degrees of freedom of the collected data. To check the impact of, if we instead limited ourselves to the lensed galaxies within the model binning range, but also to gain more insights and verify the robustness of our results, we here consider three more cases, in total four with our full range case where in the first case (Case I), we do not include the $C_{\ell s}$ obtained from photometrically observed sources in bins outside the redshifts of our parameters, in the second case (Case II), we add $C_{\ell s}$ from the higher bins, then in the third (Case III) we add all $C_{\ell s}$, which is actually our baseline in the main text, to end, in the fourth case (case IV), by including $C_{\ell s}$ from the lower outside bins but not from the higher ones.
In these scenarios, naturally Case I is expected to be the least constraining on our parameters, while in Case III would yield the strongest ones. 
We shall limit to showing the $z$ dependence error bars for all the parameters as in Fig.~\ref{fig:cases}, and only group all the $z$ and $k$ binning results in Table~\ref{tab:eta} for the $\eta$ only, being the main parameter investigated impact here. We also figure in the same table the relative error on $\eta$ when considering it as constant for all the redshift and wave-number range.

Following our original binning we observe relative errors for Case I for $E(z_i)$ within or less than 5\% and as expected tightening and reaching 1\% in Case III, with an increase of the errors with the redshift value. This confirms the previous interpretation that $E(z_i)$ is implicated in modelling the whole line of sight projection from the lensing of the sources all the way till the last observed bin. While for the $R(z_i)$ parameter, we observe for all cases, errors in the percent order, which could be explained by the fact that $R$ is essentially constrained by the spectroscopic measurements and will not be affected by missing $C_{\ell s}$. The picture is different for the $L(z_i)$ parameters, where we observe the largest difference between Case I and III, more than it was the case for $E(z_i)$ or $R(z_i)$, and that going from values in the sub-precents to reach $\sim$3\% for the last redshift bin in the most constraining case, while doubling to reach more than 400\% in the last $z$ bin in Case I, due to the fact that $L$ is essentially constrained by the projected lensed spectra from the photometric measurements with a decreasing number of lensed galaxies going up with higher redshift bins while here we additionally do not include all the $C_{\ell s}$ for bins above the last bin used for our model. The change between the cases in error values and trend for $\eta$ reflect this balance between $R$ and $L$, since the $E$ trend function of $z$ is the same for all cases, it remains that the large change in $L$ imposing its trend. Finally, we obtain by Jacobian projection, the change in the relative errors on $\eta$ as shown in Table~\ref{tab:eta_cases} where we see that we loose precision by a factor of 2 in the least constraining scheme. We also note an important observation, seen either in the plots or the tables and for the different binning scheme, that the cut of $C_{\ell s}$ from higher bins (case IV) has much more effect than when omitting those from lower bins (case II). This is due to the fact that the high redshift sources will be lensed by the intermediate ones forming the parameters of our derivation of $\eta$, while the low sources projected clustering or lensing will be only weakly affected by the change in our parameters that could occur from our interpolation method.  

\begin{table*}
\begin{centering}
\begin{tabular}{|c|c|c|}
\hline
 & low-$z$ $C_{\ell s}$ & high-$z$ $C_{\ell s}$ \\
\hline 
Case I & \xmark & \xmark \\
Case II & \xmark & \cmark \\
Case III & \cmark & \cmark \\
Case IV &  \cmark & \xmark \\
\hline 
\end{tabular}
\par\end{centering}
\begin{centering}
\caption{\label{tab:fourcases} Four cases used for the photometric Fisher analysis depending on whether we include the $C_{\ell s}$ from below (Case III and IV) or above (Case II and III) the redshift range from which we considered our model independent parameters. Case I corresponds to the one where we only restrict to galaxies within our parameters range.}
\par\end{centering}
\end{table*}

\begin{table}
\centering
\begin{tblr}{|l||c c c c||c c c c|}
\toprule
 && \SetCell[c=2]{c} Case I  & & &\SetCell[c=4]{c} Case II & \\ 
\midrule
$\sigma_\eta$ & z = 0.8 & z = 1.0 & z=1.2 & z=1.4 & z=0.8 
& z = 1.0 & z=1.2 
& z=1.4 \\ 
\midrule
   $k=0.0075$ & 0.283 & 0.222 & 0.203 &  0.608 & 0.245 & 0.195 & 0.159 & 0.115\\
   $k=0.03$ & 0.326 & 0.248 & 0.264 & 1.37 & 0.263 &  0.213  & 0.181 & 0.113\\ 
   $k=0.075$ & 0.315 & 0.206 & 0.264 & 1.90 & 0.203 & 0.173 & 0.149 & 0.098\\ 
   $k=0.125$ & 0.56 & 0.393 & 0.442 & 3.50 & 0.348  & 0.336   & 0.248 & 0.141\\ 
\midrule
   $\eta_{\rm const}$  && \SetCell[c=2]{c} 0.104  & & &\SetCell[c=4]{c} 0.057 & \\
\hline 
 && \SetCell[c=2]{c} Case III (3x2pt XC + GC$_{\rm sp}$ all obs. bins) &&& \SetCell[c=4]{c} Case IV &  \\ 
\midrule
$\sigma_\eta$ & z = 0.8 & z = 1.0 & z=1.2 & z=1.4 & z=0.8  & z = 1.0 & z=1.2 &  z=1.4 \\ 
\midrule
   $k=0.0075$ & 0.210 &  0.174 & 0.141 & 0.103 & 0.257  & 0.199 & 0.182 & 0.517\\
   $k=0.03$ & 0.226 &  0.192  & 0.162 & 0.101 & 0.304 & 0.227  & 0.242  & 1.26\\ 
   $k=0.075$ & 0.200 &  0.169 & 0.144 &  0.095 & 0.393 &  0.12 & 0.249 &  1.86\\ 
   $k=0.125$ & 0.343 &  0.326  & 0.239 & 0.137 & 0.519 &  0.382   & 0.421 & 3.45\\    
\midrule
   $\eta_{\rm const}$  && \SetCell[c=2]{c} 0.055  & & &\SetCell[c=4]{c} 0.101 & \\
\midrule
\end{tblr}
\caption{1$\sigma$ relative errors values on $\eta$, obtained from model-independent measurements at various $z$ and $k$ bins for the four cases described in Table~\ref{tab:fourcases}, along with the relative error when considering $\eta$ as constant along the whole redshift and wave-number range. 
}
\label{tab:eta_cases}
\end{table}

\section{Intrinsic alignment modeling}\label{sec:IA_model}

A simple model for the ellipticities of elliptical galaxies was proposed by \cite{Catelan:2000vm} where the intrinsic shear of the galaxy is assumed to follow the relation:
\begin{equation}
\gamma^{\rm I} = -{C \over 4\pi G}(\nabla_x^2-\nabla_y^2, 
2\nabla_x\nabla_y){\cal S}[\Psi_P],
\label{eq:model1}
\end{equation}
where $\Psi_P$ is the Newtonian potential at the time of galaxy formation, assumed to be early in the matter domination epoch, $ G$ is the Newton's gravitational constant, $x$ and $y$ are Cartesian coordinates in the plane of the sky, ${\cal S}$ is a smoothing filter that cuts off fluctuations on galactic scales, $\nabla$ is a comoving derivative and $C$ is a normalization constant that will depend in general, mainly on the luminosity of the galaxy and its other less important properties.  The original motivation for Eq.~(\ref{eq:model1}) was the assumption that halo ellipticity is perturbed by the local tidal field produced by large scale structure \citep{Catelan:2000vm}. On sufficiently large scales, the correlations in the intrinsic shear field must be determined by the large-scale potential fluctuations which, if sufficiently small, should be a linear and local function of the early potential that is then related to the linear density field via:
\begin{equation}
\Psi_P({\bf k}) = -4\pi \, \mu \, G{\bar\rho(z)\over D(z)}a^2k^{-2}\delta_{\rm lin}({\bf k}),
\end{equation}
where $\bar\rho(z)$ is the mean density of the universe, $D(z)\propto(1+z)D(z)$ is the growth factor that serves to, following \cite{Hirata:2004gc} to froze the action of the primordial field from further evolution, and $\mu$ is the function we usually introduce to account for deviation from GR in the Poisson equation. We will be interested in the weighted intrinsic shear related to the galaxy perturbation $\delta_{\rm g} = b_g\delta_{\rm lin}$ through:
\begin{equation}
\tilde\gamma^{\rm I}({\bf k}) = {C \bar\rho\over D} \, \mu \, a^2 \int 
{k_{2x}^2-k_{2y}^2 + 2k_{2x}k_{2y}\over k_2^2} \delta_{\rm lin}({\bf k}_2) \Bigl[ 
\delta^{(3)}({\bf k}_1) + {b_g\over (2\pi)^3} \delta_{\rm lin}({\bf k}_1)\Bigr] d^3{\bf k}_1,
\label{eq:gtilde}
\end{equation}
where $b_g$ is the linear galaxy bias, ${\bf k}_2\equiv{\bf k}-{\bf k}_1$ and we have chosen the wave
vector ${\bf k}$ to lie on the $x$-axis since we observe modes with ${\bf k}$ perpendicular to the line of sight.
The power spectrum of $\tilde\gamma$ would then be:
\begin{equation}
P_{\tilde\gamma^{\rm I} \tilde\gamma^{\rm I}}(k)=P^{EE}_{\tilde\gamma^{\rm I} \tilde\gamma^{\rm I}}(k) = {C^2\bar\rho^2\over D^2}\, \mu^2\,a^4 \Bigl\{ 
P_\delta^{lin}(k)
+ b_g^2\int [f_E({\bf k}_2) + f_E({\bf k}_1) ] f_E({\bf k}_2) { P_\delta^{lin}(k_1)  P_\delta^{lin}(k_2) \over (2\pi)^3}
d^3{\bf k}_1 \Bigr\},
\label{eq:eemode}
\end{equation}
where $f_E(\bf{k})$ is a geometric function that singles out correlations between the $E$-modes of the ellipticity field, while the $B$-mode correlations are zero, due to the symmetry of the tidal shear tensor.
The second term in brackets in Eq.~\ref{eq:eemode} is caused by the density weighting and is proportional to the square of the linear matter power spectrum and is sub-dominant compared to the first term on large scales when the linear alignment model is applied so that at end we arrive at:
\begin{equation}
    P_{\delta_{\rm I}\delta_{\rm I}}(k) = {C^2\bar\rho^2\over D^2}\, \mu^2\,a^4 P_{\delta_{\rm m}\delta_{\rm m}}^{lin}(k) 
\end{equation}
which we recast into 
\begin{equation}
    P_{\delta_{\rm I}\delta_{\rm I}}(k) = {\left[H_0 \, n_i(z) E(z)\right]}^2 \times \left(\,\mu\, {\cal{A}}_{\rm IA} {\cal{C}}_{\rm IA} \, \mu(k,z)\,\Omega_{{\rm m},0}\frac{{\cal{F}}_{\rm IA}(z)}{G(z,k)} \right) ^2 {P_{\delta_{\rm m}\delta_{\rm m}}^{lin}(k)} 
\end{equation}
where we recognize the quantities $W^{\rm IA}$ in the first term and $\delta_{\rm I}$ in the second term we defined respectively in Eq.~\ref{eq:iaweight} and \ref{eq:pdienla}
%%%%%%%%%%%%%%%%%%%%%%%%%%%%%%%%%%%%%%%%%%%%%%%%%%

% Don't change these lines
\bsp	% typesetting comment
\label{lastpage}
\end{document}